# Landau damping effects and evolutions of energy spread in small isochronous ring

Yingjie Li[a,*], Lanfa Wang[b], Fanglei Lin[c]

[a]*Department of Physics, Michigan State University, East Lansing, MI 48824, USA*
[b]*SLAC National Accelerator Laboratory, Menlo Park, CA 94025, USA*
[c]*Thomas Jefferson National Accelerator Facility, Newport News, VA 23606, USA*

**Abstract**

This paper presents the Landau damping effects on the microwave instability of a coasting long bunch in an isochronous ring due to finite energy spread and emittance. Our two-dimensional (2D) dispersion relation gives more accurate predictions of the microwave instability growth rates of short-wavelength perturbations than the conventional 1D formula. The long-term evolution of energy spread is also studied by measurements and simulations.

## 1. Introduction

A coasting long bunch with high beam intensity in the Small Isochronous Ring (SIR) at Michigan State University (MSU) demonstrates strong microwave instability with some unique properties that are different from the conventional ones [1-5]. For example, in the linear stage of the microwave instability, the instability growth rates are proportional to the unperturbed beam intensity $I_0$ instead of the square root of $I_0$ [2]. Pozdeyev pointed out that, in the isochronous regime, the radial coherent space charge fields of a coasting bunch with centroid wiggles may modify the slip factor, raise the working point above transition and enhance the microwave instability. This makes the instability growth rates linearly dependent on the beam intensity [3, 4]. Another unique property is that the spectral evolutions of the line charge densities are not pure exponential functions of time, instead, they are often characterized by the betatron oscillations superimposed on the exponential growth curves. These betatron oscillations are the dipole modes in the longitudinal structure of the beam due to dipole moment of the centroid offsets [6].

Pozdeyev and Bi developed their own models and theories separately to explain the mechanisms of microwave instability in the isochronous regime [3-5], respectively. Their models use the 1D (longitudinal) conventional instability growth rates formula derived exclusively for a mono-energetic and laminar beam, in which the strong radial-longitudinal coupling effects in an isochronous ring are not included completely. This may overestimate the instability growth rates, especially for the short-wavelength perturbations, because the strong Landau damping effects caused by the finite energy spread and the emittance are all neglected. Though Pozdeyev explained the suppression of the instability growth of short-wavelength perturbations by the radial-longitudinal coupling effects qualitatively [4], no quantitative discussions on the Landau damping effects are available for a coasting bunch with space charge in the isochronous regime.

To predict the microwave instability growth rates more accurately, we introduce and modify a 2D dispersion relation with Landau damping effects considering the contributions from both the finite energy spread and emittance. By doing this, it can explain the suppression of the microwave instability growth rates of the short-wavelength perturbations and predict the fastest-growing wavelength.

This paper is organized as follows. Section 2 briefly introduces the Small Isochronous Ring (SIR) at MSU and the simulation code. Section 3 discusses the limits of the conventional 1D growth rates formula, presents a modified 2D dispersion relation with Landau damping effects. Section 4 discusses the Landau damping effects in the isochronous ring by 2D dispersion relation. Section 5 presents the simulation study of microwave instability in SIR and provides benchmarking of the 2D dispersion relation with different beam parameters. Section 6 discusses the energy spread measurements and simulations for the nonlinear

* Corresponding author.
*Email address*: liyingji@msu.edu (Y. Li).

beam dynamics.

## 2. Small Isochronous ring and CYCO

The National Superconducting Cyclotron Laboratory (NSCL) at Michigan State University (MSU) constructed a low energy and low beam intensity Small Isochronous Ring (SIR) to simulate and study the space charge effects in the high power isochronous cyclotrons [1, 2] by scaling law. Its main parameters are shown in Table 1.

Table 1. Main Parameters of SIR

| Parameters | Symbol | Values |
|---|---|---|
| Ion species | | $H_2^+$ |
| Kinetic energy | $E_{k0}$ | 20 keV |
| Beam current | $I_0$ | 5-25 $\mu$A |
| Bunch length | $L_b$ | 15 cm- 5.5 m |
| Betatron tunes | $v_x$ , $v_y$ | 1.14, 1.11 |
| Bare slip factor | $\eta_0$ | $2\times10^{-4}$ |
| Circumference | $C_0$ | 6.58 m |
| Rev. period | $T_0$ | 4.77 $\mu$s |
| Life time | | ~ 200 turns |
| Beam radius | $r_0$ | ~ 0.5 cm* |
| Full chamber width | $W$ | 11.4 cm |
| Full chamber height | $H$ | 4.8 cm |

* The beam radius $r_0$ ~ 0.5 cm is calculated from the algebraic matched-beam envelope equation (4.88a) and its solution Eqs. (4.93) of Ref. [7] for a typical beam of 10 $\mu$A, 20 keV, and 50 $\pi$ mm mrad.

The Small Isochronous Ring consists of three main parts: a multi-cusp Hydrogen ion source, an injection line and a storage ring. There are four identical flat-field bending magnets in the storage ring. The pole face of each magnet is rotated by $26^{\circ}$ to provide isochronism and vertical focusing at the same time. A pair of fast-pulsed electrostatic inflectors can kick the beam with the desired length into the ring. Since no RF cavity is utilized in the storage ring, the bunch can coast in the ring with a life time of about 200 turns. There is an extraction box located in one of the drift lines of the ring, in which a pair of fast-pulsed electrostatic deflectors can kick the beam down to the Fast Faraday Cup (FFC) installed below the median ring plane. The FFC is used to monitor the longitudinal beam profiles. The energy spread measurements can also be performed if we replace the FFC assembly by an energy analyzer assembly.

A 3D Particle-In-Cell (PIC) simulation code CYCO [1] was used to study the beam dynamics with space charge in the isochronous regime. It can numerically solve the complete set of equations of motions of charged particles in a realistic 3D field map with space charge fields included. Due to the large aspect ratio between the width and height, the storage ring vacuum chamber with rectangular cross-section is simplified as a pair of infinitely large conducting plates which is parallel to the median ring plane.

## 3. 2D dispersion relation

### *3.1. Limit of 1D growth rates formula*

The conventional 1D growth rates formula for the microwave instability of a mono-energetic and laminar beam used in Ref. [4] is:

$$\tau^{-1}(k) = \omega_0 \sqrt{-i\frac{\eta e I_0 k R Z(k)}{2\pi\beta^2 E}}, \tag{1}$$

where $\omega_0$ is the angular revolution frequency of the on-momentum particles, $\eta = \alpha - 1/\gamma^2$ is the slip factor, $\alpha$ is the momentum compaction factor, $\gamma$ is the relativistic energy factor of the on-momentum particle, $e$ is the electron charge, $I_0$ is the unperturbed beam intensity, $k$ is the perturbation wavenumber of the longitudinal charge density, $R$ is the average ring radius, $Z(k)$ is the longitudinal space charge (LSC) impedance, $\beta$ is the relativistic speed factor, $E$ is the total energy of the on-momentum charged particle. Essentially, the 1D dispersion relation Eq. (23) of Ref. [5] is the same as the 1D growth rates formula Eq. (1), if we express the longitudinal electric field by the LSC impedance.

If the transverse dimension of a vacuum chamber is much greater than the beam diameter, the image charge effects of the chamber wall can be neglected in the short wavelength limit [8-10], therefore the longitudinal monopole mode space charge impedance $Z(k)$ of a circular beam can be approximated from the on-axis space charge field as [4]:

$$Z(k) = Z_{0,sc}^{\parallel}(k) = i\frac{2Z_0 R}{k\beta {r_0}^2}[1 - \frac{kr_0}{\gamma}K_1(\frac{kr_0}{\gamma})], \tag{2}$$

where $Z_0$ = 377 Ohm ($\Omega$) is the impedance of free space, $r_0$ is the beam radius, $K_1(x)$ is the modified Bessel function of the second kind.

For a coasting long bunch with strong space charge effects in the isochronous ring, the longitudinal space charge fields may induce the coherent energy deviations and the associated radial offsets of the local centroids. Consequently, there are centroid wiggles along the bunch. Assume the longitudinal distribution of the radial centroid offsets is a sinusoidal function of the longitudinal coordinate $z$ with a wavenumber $k_c$, in the first-order approximation, we can choose $k \approx k_c$ and use the same $k$ in the expressions of $\eta(k)$ and $Z(k)$ in Eq. (1) just as treated in Ref. [4] (please check Eqs. (2), (12), (13), and (14) in Ref. [4]). Then $\eta$ may be approximated by the space-charge modified coherent slip factor $\eta_{sc}(k)$ as [3, 4]

$$\eta \approx \eta_{sc}(k) = \frac{eI_0}{2\pi\varepsilon_0 \gamma m_{H_2^+} \omega_0^3 r_0^2 R}[1 - \frac{kr_0}{\gamma}K_1(\frac{kr_0}{\gamma})], \tag{3}$$

where $m_{H_2^+}$ is the rest mass of $H_2^+$ ion, $\varepsilon_0 = 8.85\times10^{-12}$ F m$^{-1}$ is the permittivity of free space. Note the relativistic factor $\gamma$ is introduced in Eq. (3) to make the original expressions of $\eta_{sc}(k)$ in Refs. [3, 4] compatible with the high energy beams.

The LSC impedance in Eq. (2) is evaluated from the 1D space charge field model on beam axis. While Venturini [11] pointed out that the granularity of the beam distribution in transverse plane may induce the transverse field fluctuations (shot noise). The transverse field fluctuations may translate into the longitudinal fluctuations of the electric field making the 1D field model invalid when perturbation wavelength is smaller than $2\pi r_0/\gamma$ or $kr_0/\gamma$ >0.5. In addition, the off-axis LSC field is always weaker than the on-axis LSC field within the beam radius. Both of these 3D effects should be taken into account in the accurate evaluation of the LSC impedance. Ratner [12] studied the above 3D space charge effects and found that, instead of the on-axis LSC field, if the mean LSC field averaged over the cross-section of a round beam with uniform transverse density is used, the 1D and 3D field models give nearly identical LSC fields. The formula for the average LSC impedance is [13]

$$Z(k) = Z_{0,sc}^{\parallel}(k) = i\frac{2Z_0 R}{k\beta {r_0}^2}[1 - 2I_1(\frac{kr_0}{\gamma})K_1(\frac{kr_0}{\gamma})], \tag{4}$$

where $I_1(x)$ is the modified Bessel function of the first kind. It is believed to be a more accurate description of the collective LSC effects than the conventional one in Eq. (2) in the short-wavelength limits.

Ref. [4] uses the following formalism to derive the space-charge modified coherent slip factor in Eq. (3): the space charge field induces the centroid wiggles, there will be coherent radial space charge field $E_{x,sc}$, it produces positive increments in the dispersion function *D*, the momentum compaction factor $\alpha$, and the coherent slip factor $\eta_{sc}(k)$, finally, the space-charge modified coherent slip factor $\eta_{sc}(k)$ can be determined, i.e., $E_{x,sc}\rightarrow\Delta D_{sc}\rightarrow\Delta\alpha_{sc}\rightarrow\Delta\eta_{sc}(k)\rightarrow\eta_{sc}(k)$. But this procedure is not accurate enough for a beam with finite coherent energy deviation and emittance in an isochronous ring. In Ref. [14], the transformation of the longitudinal coordinate *z* with respect to the bunch center is

$$z = z_0 + R_{51}x_0 + R_{52}x_0^{'} + R_{56}\delta, \tag{5}$$

where $x_0$ and $x_0' = dx_0/ds$ are the radial coordinate and velocity slope at $s$ = 0, respectively, $\delta = \Delta p/p$ is the fractional *momentum* deviation (Note in Ref. [14], $\delta$ was defined differently as the fractional *energy* deviation $\delta=\Delta E/E$ of an ultra-relativistic electron particle with $\beta\approx1$, since $\delta=\Delta p/p\approx(\Delta E/E)/\beta^2\approx\Delta E/E$), $R_{51}(s)$, $R_{52}(s)$, and $R_{56}(s)$ are the transfer matrix elements and depend on the path length *s* with respect to a reference point of storage ring. For a coasting beam with space charge in the isochronous ring, the space–charge modified parameters of the slip factor $\eta_{sc}$, the momentum compaction factor $\alpha_{sc}$, the transition gamma $\gamma_{t,sc}$, the element $R_{56,sc}$, and the local dispersion function $D_{sc}(s)$ are related to each other by:

$$\eta_{sc} = \alpha_{sc} - \frac{1}{\gamma^2} = \frac{1}{\gamma_{t,sc}^2} - \frac{1}{\gamma^2} = -\frac{R_{56,sc}(C_0)}{C_0}, \tag{6}$$

$$\alpha_{sc} = \frac{1}{\gamma_{t,sc}^2} = -\frac{R_{56,sc}(C_0)}{C_0} + \frac{1}{\gamma^2} = \frac{1}{C_0}\int_L \frac{D_{sc}(s)}{\rho(s)}ds = <\frac{D_{sc}(s)}{\rho(s)}>, \tag{7}$$

where $\rho(s)$ is the local radius of curvature of trajectory, <…> denotes the average value over the ring circumference $C_0$. From Eqs. (6)(7) and the formalism used in Ref. [4], we can see that only the contribution of the momentum compaction factor $\alpha_{sc}$ or the element $R_{56,sc}$ is considered in the modification of $\eta_{sc}(k)$. In Ref. [14], the parameters $x_0$, $x_0'$, and $z_0$, at $s$=0 are regarded as constants of motion, they are related to the current coordinates of the particle $x$, $x'$ and $z$ at position $s$ by a canonical transformation

$$\begin{aligned}
x_0(x,x',\delta,s) &= \sqrt{\frac{\hat\beta_0}{\hat\beta}}(x-D\delta)\cos\psi - \sqrt{\hat\beta_0\hat\beta}[x'-D'\delta+\frac{\hat\alpha}{\hat\beta}(x-D\delta)]\sin\psi,\\
x_0'(x,x',\delta,s) &= \frac{x-D\delta}{\sqrt{\hat\beta\hat\beta_0}}\sin\psi + \sqrt{\frac{\hat\beta}{\hat\beta_0}}[x'-D'\delta+\frac{\hat\alpha}{\hat\beta}(x-D\delta)]\cos\psi,\\
z_0(x,x',\delta,s) &= z - R_{56}\delta - x_0R_{51} - x_0'R_{52},
\end{aligned} \tag{8}$$

where $\hat\alpha$, $\hat\beta_0$, and $\hat\beta$ are the Courant-Snyder parameters, $\psi$ is the phase advance. Their derivatives with respect to $\delta$ are

$$\begin{aligned}
\frac{\partial x_0}{\partial\delta} &= -\sqrt{\frac{\hat\beta_0}{\hat\beta}}D\cos\psi + \sqrt{\hat\beta_0\hat\beta}(D'+\frac{\hat\alpha}{\hat\beta}D)\sin\psi,\\
\frac{\partial x_0'}{\partial\delta} &= -\frac{D}{\sqrt{\hat\beta\hat\beta_0}}\sin\psi - \sqrt{\frac{\hat\beta}{\hat\beta_0}}(D'+\frac{\hat\alpha}{\hat\beta}D)\cos\psi,\\
\frac{\partial z_0}{\partial\delta} &= -R_{56} - \frac{\partial x_0}{\partial\delta}R_{51} - \frac{\partial x_0'}{\partial\delta}R_{52},
\end{aligned} \tag{9}$$

which usually are non-zero parameters. Eq. (5) show $\Delta z = z - z_0$ is determined by $R_{51}$, $R_{52}$, and $R_{56}$. The ring is isochronous if $\Delta z$ = 0 after one revolution. Accordingly, the exact expression of the slip factor at *s* can be calculated as $\eta(s,k)=-(d\Delta z/d\delta)/C_0=-[R_{51}(s,s+C_0)\partial x_0/\partial\delta + R_{52}(s,s+C_0)\partial x_0'/\partial\delta + R_{56}(s,s+C_0)]/C_0$, and it also depends on $R_{51}(s,s+C_0)$, $R_{52}(s,s+C_0)$ if $\partial x_0/\partial\delta\neq0$ and $\partial x_0'/\partial\delta\neq0$. Here $R_{51}(s,s+C_0)$, $R_{52}(s,s+C_0)$ and $R_{56}(s,s+C_0)$ are the transfer matrix elements between *s* and $s+C_0$. The space–charge modified slip factor expressed in Eq. (6) is only a special case at $s$=0, $\partial x_0/\partial\delta$ =0, and $\partial x_0'/\partial\delta$=0.

### *3.2. Space-charge modified tunes and transition gamma in the isochronous regime*

The radial space charge fields may modify the radial tunes and transition gamma in the isochronous regime [3-5]. Due to the large ratios between the full chamber width (~11.4 cm), full chamber gap (~4.8cm) and the beam diameter (~1 cm), the image charge effects caused by the vacuum chamber are small for perturbation wavelength less than 5 cm [8-10]. Then Pozdeyev's model [3, 4] of a uniform circular beam with centroid wiggling in free space can be used to calculate the radial space charge fields and modified tunes. Assuming the total radial offset of a particle is $x = x_c + x_\beta$, where $x_c$ is the beam centroid offset $x_c = a_c \cos(kz)$, $k$ is the wave number of radial offset perturbations of local beam centroids with respect to design orbit along $z$, $x_\beta$ is the radial offset of single particle due to the betatron oscillation. The equations of coherent and incoherent radial motions can be expressed as:

$$x_c'' + \frac{\nu_x^2}{R^2} x_c = \frac{\delta_{coh}}{R} + \frac{eE_{x,coh}}{\gamma m_{H_2^+} \beta^2 c^2}, \tag{10}$$

$$x_\beta'' + \frac{\nu_x^2}{R^2} x_\beta = \frac{\delta_{inc}}{R} + \frac{eE_{x,inc}}{\gamma m_{H_2^+} \beta^2 c^2}, \tag{11}$$

where $\nu_x$ is the bare radial betatron tune, $\delta_{coh}$ and $\delta_{inc}$ are the coherent and incoherent fractional momentum deviations, $E_{x,coh}$ and $E_{x,inc}$ are the coherent and incoherent radial space charge fields [4, 5], respectively, and can be expressed as

$$E_{x,coh} = \xi_{coh} \gamma m_{H_2^+} \omega_0^2 x_c / e, \quad E_{x,inc} = \xi_{inc} \gamma m_{H_2^+} \omega_0^2 x_\beta / e. \tag{12}$$

Here

$$\xi_{coh} = \frac{eI_0}{2\pi\varepsilon_0 \gamma m_{H_2^+} \omega_0^3 r_0^2 R}[1 - \frac{kr_0}{\gamma} K_1(\frac{kr_0}{\gamma})], \quad \xi_{inc} = \frac{eI_0}{2\pi\varepsilon_0 \gamma m_{H_2^+} \omega_0^3 r_0^2 R}, \tag{13}$$

are two unitless parameters. For a typical SIR beam with $0 < \xi_{coh} << 1$ and $0 < \xi_{inc} << 1$, the coherent and incoherent radial tunes can be easily obtained from Eqs. (10) - (13) as

$$\nu_{x,coh} \approx \nu_x (1 - \frac{\xi_{coh}}{2\nu_x^2}), \quad \nu_{x,inc} \approx \nu_x (1 - \frac{\xi_{inc}}{2\nu_x^2}). \tag{14}$$

Here the coherent radial tune $\nu_{x,coh}$ and incoherent radial tune $\nu_{x,inc}$ stand for the number of betatron oscillations per revolution of a local centroid and a single particle, respectively. According to Ref. [5], the space-charge modified coherent and incoherent transition gammas in an isochronous accelerator are

$$\gamma_{t,coh}^2 = \frac{\delta p / p}{\delta R / R} = 1 - n - \xi_{coh}, \qquad \gamma_{t,inc}^2 = \frac{\delta p / p}{\delta R / R} = 1 - n - \xi_{inc}. \tag{15}$$

where $n = -(r/B)(\partial B/\partial r)$ is the magnetic field index. For the SIR with $n < 0$, $|n| << 1$, if the space charge effects are negligible (i.e., $\xi_{coh} = \xi_{inc} = 0$, $\gamma_{t,0}^2 = 1 - n$), the bare slip factor is $\eta_0 = 1/\gamma_{t,0}^2 - 1/\gamma^2 \approx 1 + n - 1/\gamma^2 \approx 2 \times 10^{-4}$. Then

$$\frac{1}{\gamma_{t,coh}^2} - \frac{1}{\gamma^2} \approx \eta_0 + \xi_{coh}, \qquad \frac{1}{\gamma_{t,inc}^2} - \frac{1}{\gamma^2} \approx \eta_0 + \xi_{inc}. \tag{16}$$

### *3.3. 2D dispersion relation*

For a hot beam with large energy spread and emittance in an isochronous ring, the Landau damping effects are important due to the strong radial-longitudinal coupling. Hence a multi-dimensional dispersion relation including both the longitudinal and radial dynamics is needed. Usually the vertical motions of particles can be regarded as decoupled from their radial and longitudinal motions. In this section, first, we would like to summarize and comment the main procedures and definitions used in Ref. [14], where a 2D (longitudinal and radial) dispersion relation was derived for the coherent synchrotron radiation (CSR) of an ultra-relativistic electron beam in a conventional storage ring. Based on this model, we can derive a 2D dispersion relation for the microwave instability of the non-relativistic $H_2^+$ beam in an isochronous ring.

### *3.3.1 Review of the 2D dispersion relation for CSR instability of ultra-relativistic electron beams in non-isochronous regime*

First, Ref. [14] defined a 2D Gaussian beam model with an initial equilibrium beam distribution function

$$f_0 = \frac{n_b}{2\pi\varepsilon_{x,0}} \exp[-\frac{x_0^2 + (\hat{\beta}_0 x_0')^2}{2\varepsilon_{x,0}\hat{\beta}_0}] g(\delta + \hat{u} z_0), \tag{17}$$

where
$$g(\delta) = \frac{1}{\sqrt{2\pi}\sigma_\delta} \exp(-\frac{\delta^2}{2\sigma_\delta^2}), \tag{18}$$

$n_b$ is the linear number density of the beam, $\varepsilon_{x,0}$ is the initial radial emittance, $x_0$ is the initial radial offset, $x_0' = dx_0/ds$ is the initial radial velocity slope, $\hat{\beta}_0$ is the betatron function at $s = 0$, $\delta$ is the uncorrelated fractional *momentum* deviation, $\hat{u}$ is the chirp parameter which accounts for the correlation between the longitudinal position $z$ of the particle in the bunch and its momentum deviation $\delta$, $\sigma_\delta$ is the *uncorrelated* fractional RMS *momentum* spread (Note in Ref. [14], $\sigma_\delta$ was defined differently as the *uncorrelated* fractional RMS *energy* spread for an ultra-relativistic electron beam with $\beta \approx 1$). Then the perturbed distribution function $f_1$ is assumed to have a sinusoidal dependence on $z_0$ as

$$f_1(x_0, x_0', z_0, \delta_0, s) = f_k(x_0, x_0', \delta_0, s) e^{ikz_0}, \tag{19}$$

where $\delta_0 = \delta + \hat{u} z_0$ is the total fractional momentum deviation including both the *uncorrelated* and *correlated* fractional momentum deviation. Plugging the distribution function of $f=f_0+f_1$ into the linearized Vlasov equation, after lengthy derivations, a Volterra integral equation is derived as

$$g_k(s) = g_k^{(0)}(s) + \int_0^s K(s', s) g_k(s') ds', \tag{20}$$

where

$$g_k(s) = \int dx_0 dx_0' d\delta_0 f_k(x_0, x_0', \delta_0, s) \exp\{-ikC(s)[\delta_0 R_{56}(s) + x_0 R_{51}(s) + x_0' R_{52}(s)]\}, \tag{21}$$

$$g_k^{(0)}(s) = \int dx_0 dx_0' d\delta_0 f_k(x_0, x_0', \delta_0, 0) \exp\{-ikC(s)[\delta_0 R_{56}(s) + x_0 R_{51}(s) + x_0' R_{52}(s)]\}, \tag{22}$$

$$C(s) = \frac{1}{1 - \hat{u} R_{56}(s)}, \tag{23}$$

is the bunch length compression factor, $K(s', s)$ is the kernel of integration. The perturbed harmonic line density with wave number $k$ at $(z, s)$ is

$$n_{1,k}(z, s) = \int dx_0 dx_0' d\delta_0 f_1 = C(s) g_k(s) e^{-ikC(s)z}, \tag{24}$$

we can see that $|C(s)g_k(s)|$ is just the amplitude of the perturbed line density at $s$. For storage rings, the linear chirp factor $\hat{u} = 0$ , the compression factor $C(s)$=1,by smooth approximations of $\varepsilon_{x,0} = \sigma_x^2 \nu_x / R$ , $\hat{\beta} = R/\nu_x$, $\psi = \nu_x s / R$, $D = R/\nu_x^2$, $\hat{\alpha} = 0$, $D' = 0$, the integral kernel in Eq. (20) is simplified as

$$K(s', s) = K_1(\chi) = -\frac{ie\Lambda_0}{\nu_x^2 \gamma m_{e^-} c C_0} k Z(k) [-\frac{R}{\nu_x} \sin(\frac{\nu_x \chi}{R}) + \chi] e^{-(k\sigma_x/\nu_x)^2 [1 - \cos(\nu_x \chi / R)] - (k\sigma_\delta / \nu_x^2)^2 \chi^2 / 2}, \tag{25}$$

where $\Lambda_0 = en_b$ is the unperturbed line charge density, $m_{e^-}$ is the rest mass of electron, $Z(k)$ is the CSR impedance in unit of Ohm (Ω), $\sigma_x$ is the RMS beam radius, $\chi = s - s'$ is the relative path length difference between two positions at $s$ and $s'$, specifically, if we choose $s'=0$, then $\chi = s$. Note that the Eq. (25) uses the SI instead of CGS system of units as in Ref. [14]. By smooth approximation, the kernel $K(s', s)$ is only dependent on the parameter $\chi=s - s'$ and radial tune $\nu_x$. Applying Laplace transform to the two sides of Eq. (20) yields an algebraic equation

$$\hat{g}_k(\mu) = \frac{\hat{g}_k^{(0)}(\mu)}{1-\hat{K}(\mu)}, \tag{26}$$

where $\mu$ is the complex Laplace variable and

$$\hat{g}_k(\mu) = \int_0^\infty ds g_k(s) e^{-\mu s}, \tag{27}$$

$$\hat{g}_k^{(0)}(\mu) = \int_0^\infty ds g_k^{(0)}(s) e^{-\mu s}, \tag{28}$$

$$\hat{K}(\mu) = \int_0^\infty d\chi K_1(\chi) e^{-\mu\chi}, \tag{29}$$

are the Laplace images of $g_k(s)$, $g_k^{(0)}(s)$ and $K_1(\chi)$, respectively. The relation of $1-\hat{K}(\mu)=0$ for the denominator of Eq. (26) determines the dispersion relation. Finally the 2D dispersion relation for the CSR instability of an ultra-relativistic electron beam in a non-isochronous storage ring is derived in Appendix B of Ref. [14] as

$$1 = -\frac{ie\Lambda_0}{\nu_x^{\,2}\gamma m_{e^-} c C_0} kZ(k) \int_0^\infty d\chi e^{-\mu\chi} [\chi - \frac{R}{\nu_x}\sin\frac{\nu_x \chi}{R}] e^{-(k\sigma_x/\nu_x)^2[1-\cos(\nu_x\chi/R)]-(k\sigma_E/E\nu_x^{\,2})^2\chi^2/2}, \tag{30}$$

Note $\sigma_\delta \approx \sigma_E/E$ has been used in Eq. (30), where $\sigma_E$ is the RMS energy spread, and the SI system of units is used in Eq. (30). Eq. (30) is an integral equation which determines the relations between the wavenumber $k$ and the complex Laplace variable $\mu$. For a fixed $k$, the values of $\mu$ can be solved numerically.

Ref. [14] did not explicitly interpret the CSR instability growth rates from the solutions to Eq. (30). Theoretically speaking, $g_k(s)$ can be calculated by inverse Laplace transform (Fourier-Mellin transform)

$$g_k(s) = \frac{1}{2\pi i}\int_{\sigma-i\infty}^{\sigma+i\infty} d\mu \frac{\hat{g}_k^{(0)}(\mu)}{1-\hat{K}(\mu)} e^{\mu s}, \tag{31}$$

where $\sigma$ is a positive real number. The integration is along the Bromwich contour, which is a line parallel to the imaginary $\mu$-axis and to the right of all the singularities satisfying $1-\hat{K}(\mu)=0$ in the complex $\mu$-plane. In practice, the integral in Eq. (31) poses a great difficulty in mathematics due to complexity of the integrand. A popular method dealing with this difficulty is widely used in the Plasma Physics [15-18] applying Cauchy's residue theory to an equivalent Bromwich contour. First, the Bromwich contour is deformed by analytic continuation, and then the solutions of $g_k(s)$ can be evaluated by the residues of the poles using Cauchy's residue theorem as

$$g_k(s) \approx \sum_j Rsd[\hat{g}_k(\mu)e^{\mu s}, \mu_j] = \sum_j \lim_{\mu\to u_j}[(\mu-\mu_j)\hat{g}_k(\mu)e^{\mu s}] = \sum_j e^{\mu_j s} Rsd[\hat{g}_k(\mu), \mu_j], \tag{32}$$

where $Rsd[\hat{g}_k(\mu), \mu_j]$ stands for the residue of $\hat{g}_k(\mu)$ at the pole $\mu_j$. Using the relation $s = \beta ct$, where $t$ is the time, the temporal evolution of $g_k(s)$ becomes

$$g_k(t) \approx \sum_j e^{\mu_j \beta ct} Rsd[\hat{g}_k(\mu), \mu_j]. \tag{33}$$

For a storage ring, $C(s)$ =1, Eq. (24) gives the amplitude of perturbed harmonic line density with wave number $k$ as

$$| n_{1,k}(z,t) | = | g_k(t) |. \tag{34}$$

Eqs. (33)(34) show that, for a pole at $\mu_j$, (a) If $Re(\mu_j) < 0$, the *k-th* Fourier component of the line density damps exponentially at a rate of $\tau^{-1} = Re(\mu_j)\beta c$; (b) If $Re(\mu_j) > 0$, this pole may induce the CSR instability which grows exponentially at a rate of $\tau^{-1} = Re(\mu_j)\beta c$. The total instability growth rates are dominated by the pole $\mu_j$ which has the greatest positive real part.

### *3.3.2 2D dispersion relation for microwave instability of low energy beam in isochronous regime*

The 2D dispersion relation Eq. (30) can be modified to study the space-charge induced microwave instability of a low energy coasting $H_2^+$ bunch in the SIR. In the derivation of Eq. (30), the term which is proportional to $1/\gamma^2$ is neglected in the longitudinal equation of motion due to $\gamma >> 1$. In addition, in Eq. (30), the method of smooth approximation is used to express all the beam optics parameters, such as the betatron function, phase advance, dispersion function, $R_{52}$, $R_{52}$, and $R_{56}$ as functions of radial tune $\nu_x$. Because the space charge effects are also neglected, the radial betatron tune $\nu_x$ in Eq. (30) is a $k$-independent constant. While for a coasting beam with space charge in the SIR, the space charge fields may modify the radial tunes and beam optics parameters. These neglected terms and space charge effects should be considered in the 2D dispersion relation for the SIR beam. Hence Eq. (1) and Eq. (4) of Ref. [14] should be modified as

$$\frac{dz}{ds} = -\frac{x}{\rho(s)} \rightarrow -\frac{x}{\rho(s)} + \frac{\delta}{\gamma^2}, \tag{35}$$

$$R_{56}(s) = -\int_0^s \frac{D(s')}{\rho(s')} ds' \rightarrow -\int_0^s \frac{D_{sc}(s')}{\rho(s')} ds' + \frac{s}{\gamma^2}. \tag{36}$$

Consequently, using relative path length difference $\chi = s - s'$, the increment of $R_{56}$ from $s'$ to $s$ in Eq. (B4) of Ref. [14] should be modified as

$$\Delta R_{56}(s', s) = -\frac{1}{\nu_x^2}\chi \rightarrow -(\frac{1}{\gamma_{t,sc}^2} - \frac{1}{\gamma^2})\chi. \tag{37}$$

When the elements $R_{51}$ and $R_{52}$ are included, the corresponding modified increment of $R_{56}$ becomes

$$\Delta \tilde{R}_{56}(s', s) = -\frac{1}{\nu_x^2}[\chi - \frac{R}{\nu_x}\sin(\frac{\nu_x \chi}{R})] \rightarrow -(\frac{1}{\gamma_{t,sc}^2} - \frac{1}{\gamma^2})\chi + \frac{R}{\nu_{x,sc}^3}\sin(\frac{\nu_{x,sc}\chi}{R}). \tag{38}$$

where $\nu_{x,sc}$ is the space-charge modified radial tune.

Note that :

(a) In Eq. (37), for the longitudinal dynamics in the isochronous regime, we cannot use the method of smooth approximation to express $\Delta R_{56}(s', s)$ by $-(1/\nu_{x,sc}^2 - 1/\gamma^2)\chi$ directly due to smallness of the slip factor, otherwise it will induce considerable errors. We may use $\Delta R_{56}(s', s) = -(1/\gamma_{t,sc}^2 - 1/\gamma^2)\chi$ instead. While the sinusoidal function term in Eq. (38) is contributed from $R_{51}$ and $R_{52}$, and it can be estimated as a function of the radial betatron tune $\nu_{x,sc}$ using the smooth approximation.

(b) In Eqs. (36)-(38), $R_{56}(s) = \partial z/\partial\delta$ is the linear correlation coefficient between the longitudinal coordinate $z$ at $s$ and the fractional momentum deviation $\delta$ at $s=0$. $\Delta R_{56}(s', s)$ is the increment of $R_{56}$ between $s'$ and $s$ without the effects of $R_{51}$ and $R_{52}$

$$\Delta R_{56}(s', s) = R_{56}(s) - R_{56}(s'). \tag{39}$$

$\Delta\tilde{R}_{56}(s', s)$ differs from $R_{56}(s)$ and $\Delta R_{56}(s', s)$ by including the effects of $R_{51}$ and $R_{52}$. According to Appendix A of Ref. [14], for a coasting beam in the SIR, $\Delta\tilde{R}_{56}(s', s)$ can be simplified as

$$\Delta\tilde{R}_{56}(s', s) = \Delta R_{56}(s', s)|_{s'} + \Delta R_{51}(s', s)\frac{\partial x_0}{\partial\delta}|_{s'} + \Delta R_{52}(s', s)\frac{\partial x_0'}{\partial\delta}|_{s'}. \tag{40}$$

Now we can substitute Eqs. (37) and (38) into Eq. (30) to get the 2D dispersion relation for the SIR beam. In the substitution, in the square bracket of the integrand between the two exponential functions of Eq. (30), the space-charge modified transition gamma $\gamma_{t,sc}$ and the radial tune $\nu_{x,sc}$ should be replaced by the coherent ones of $\gamma_{t,coh}$ and $\nu_{x,coh}$, respectively. While $\gamma_{t,sc}$ and $\nu_{x,sc}$ in the last exponential function of Eq. (30) should be replaced by the incoherent ones of $\gamma_{t,inc}$ and $\nu_{x,inc}$, respectively. If the uncorrelated fractional RMS *momentum* spread $\sigma_\delta$ is replaced by the RMS *energy* spread $\sigma_E$ using the relation $\sigma_\delta=\sigma_E/(\beta^2 E)$, where $E$ is the total energy of the on-momentum particle, finally, the 2D dispersion relation for a low energy SIR beam in the SI system of units becomes

$$1=-\frac{ie\Lambda_0}{\beta\gamma m_{H_2+}cC_0}kZ_{0,sc}^{\parallel}(k)\int_0^\infty d\chi e^{-\mu\chi}[\left(\frac{1}{\gamma^2_{t,coh}}-\frac{1}{\gamma^2}\right)\chi-\frac{R}{\nu^3_{x,coh}}\sin\frac{\nu_{x,coh}\chi}{R}]$$
$$e^{-(k\sigma_x/\nu_{x,inc})^2[1-\cos(\nu_{x,inc}\chi/R)]-\frac{1}{2}[\frac{k\sigma_E}{\beta^2 E}(1/\gamma^2_{t,inc}-1/\gamma^2)\chi]^2}. \quad (41)$$

Note that the 2D dispersion relation Eq. (41) is derived for a Gaussian beam model without the coherent radial centroid offsets and energy deviations. Therefore it is only valid for predictions of the long-term microwave instability growth rates in an isochronous ring neglecting the line charge density oscillations due to dipole moments of the centroid offsets. Here the term 'long-term' stands for multi-periods of betatron oscillations in the time scale. When the dispersion relation Eq. (41) is to be solved numerically, a large but finite real number can be set as the upper limit of $\chi$ instead of infinity to calculate the integral.

## 4. Landau damping effects in isochronous ring

### *4.1. Space-charge modified coherent slip factors*

For the SIR beam with typical beam intensities, usually $|\eta_0| << \xi_{coh}$, when the space charge effects are considered. Then in the first-order approximation, according to Eq. (16), the space-charge modified coherent slip factor without the effects of $R_{51}$ and $R_{52}$ may be estimated as

$$\eta_{coh}=\eta_{R56}=\frac{1}{\gamma^2_{t,coh}}-\frac{1}{\gamma^2}\approx\eta_0+\xi_{coh}\approx\xi_{coh}, \quad (42)$$

which is essentially the same as Eq. (12) in Ref. [4].

For a ring lattice with average radius $R$ and space-charge modified radial tune $\nu_{x,sc}$, by smooth approximation of $\hat{\beta}=R/\nu_x$, $\psi=\nu_x s/R$, $D=R/\nu_x^2$, $\hat{\alpha}=0$, $D'=0$, the increments of $R_{51}$ and $R_{52}$ between $s'$ and $s$ can be calculated from Eq. (B2) and Eq. (4) of Ref. [14] as

$$\Delta R_{51}(s',s)=-\frac{1}{\nu_{x,sc}}[\sin(\frac{\nu_{x,sc}}{R}s)-\sin(\frac{\nu_{x,sc}}{R}s')], \quad (43)$$

$$\Delta R_{52}(s',s)=\frac{1}{\nu^2_{x,sc}}[\cos(\frac{\nu_{x,sc}}{R}s)-\cos(\frac{\nu_{x,sc}}{R}s')], \quad (44)$$

According to Eq. (9) (i.e., Eq. (20) of Ref. [14]),

$$\frac{\partial x_0}{\partial\delta}|_{s'}=-\frac{R}{\nu^2_{x,sc}}\cos(\frac{\nu_{x,sc}}{R}s'), \quad (45)$$

$$\frac{\partial x_0'}{\partial \delta}|_{s'} = -\frac{1}{\nu_{x,sc}}\sin(\frac{\nu_{x,sc}}{R}s'), \tag{46}$$

Then by Eqs. (40)(43)(44)(45)(46), in the second-order approximation, taking into account the contributions from the matrix elements $R_{51}$ and $R_{52}$ to the longitudinal beam dynamics, the space-charge modified coherent slip factor can be calculated as

$$\tilde{\eta}_{coh} = -\frac{\Delta\tilde{R}_{56}(s,s+C_0)}{C_0} = -[\Delta R_{56}(s,s+C_0)|_s + \Delta R_{51}(s,s+C_0)\frac{\partial x_0}{\partial\delta}|_s + \Delta R_{52}(s,s+C_0)\frac{\partial x_0'}{\partial\delta}|_s]/C_0$$

$$= \eta_{R56} + \eta_{R51}(s) + \eta_{R52}(s) = \eta_{R56} - \frac{1}{2\pi\nu^3_{x,coh}}\sin(2\pi\nu_{x,coh}), \tag{47}$$

where

$$\eta_{R51}(s) = -\Delta R_{51}(s,s+C_0)\frac{\partial x_0}{\partial\delta}|_s / C_0 = -\frac{1}{2\pi\nu^3_{x,coh}}[\sin(\frac{\nu_{x,coh}}{R}(s+C_0)) - \sin(\frac{\nu_{x,coh}}{R}s)]\cos(\frac{\nu_{x,coh}}{R}s), \tag{48}$$

$$\eta_{R52}(s) = -\Delta R_{51}(s,s+C_0)\frac{\partial x_0'}{\partial\delta}|_s / C_0 = \frac{1}{2\pi\nu^3_{x,coh}}[\cos(\frac{\nu_{x,coh}}{R}(s+C_0)) - \cos(\frac{\nu_{x,coh}}{R}s)]\sin(\frac{\nu_{x,coh}}{R}s), \tag{49}$$

$$\eta_{R56} = -\Delta R_{56}(s,s+C_0)/C_0 = \frac{1}{\gamma^2_{t,sc}} - \frac{1}{\gamma^2}, \tag{50}$$

are slip factors contributed from the matrix elements $R_{51}$, $R_{52}$ and $R_{56}$, respectively. We can see $\eta_{R51}$ and $\eta_{R52}$ are functions of $s$, while $\eta_{R56}$ is independent of $s$. The last term of Eq. (47) is contributed from the combined effects of $R_{51}$ and $R_{52}$ and is independent of $s$.

Assuming a SIR bunch has beam intensity $I_0 = 1.0\ \mu$A, kinetic Energy $E_{k0} = 19.9$ keV, radial and vertical emittance $\varepsilon_{x,0} = \varepsilon_{y,0} = 50\ \pi$ mm mrad, the calculated slip factors by Eqs. (47)-(50) as functions of line charge perturbation wavelength $\lambda$ at $s = C_0$ and $s = 10\ C_0$ are shown in Fig. 1. If we increase the beam intensity to 10 $\mu$A, the calculated slip factors at $s = C_0$ and $s = 10\ C_0$ are shown in Fig. 2.

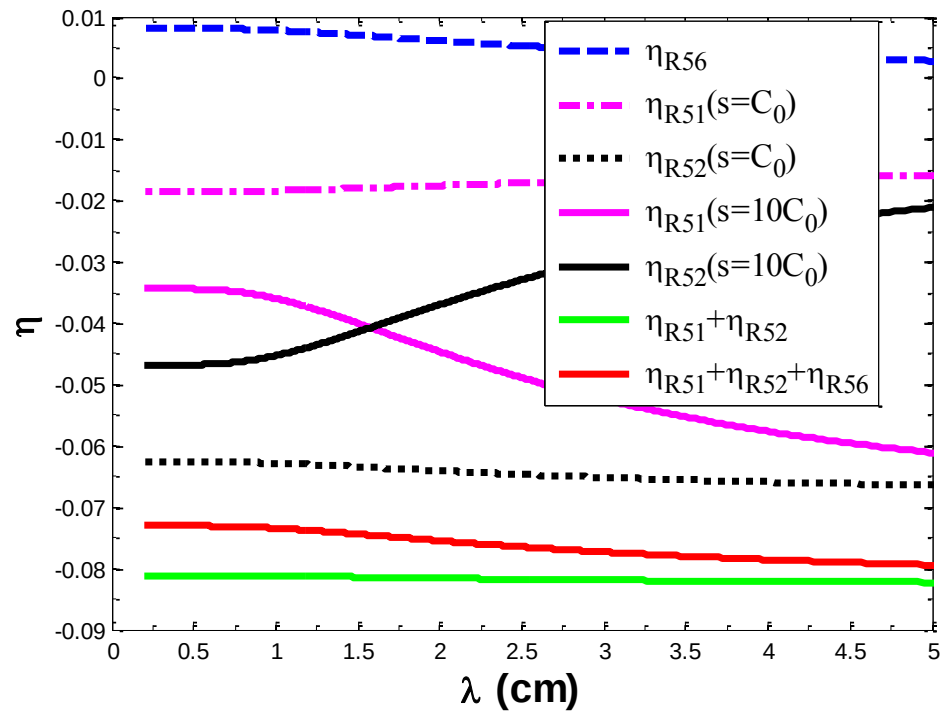

Fig. 1. Slip factors for $I_0 = 1.0\ \mu$A at $s = C_0$ and $s = 10\ C_0$.

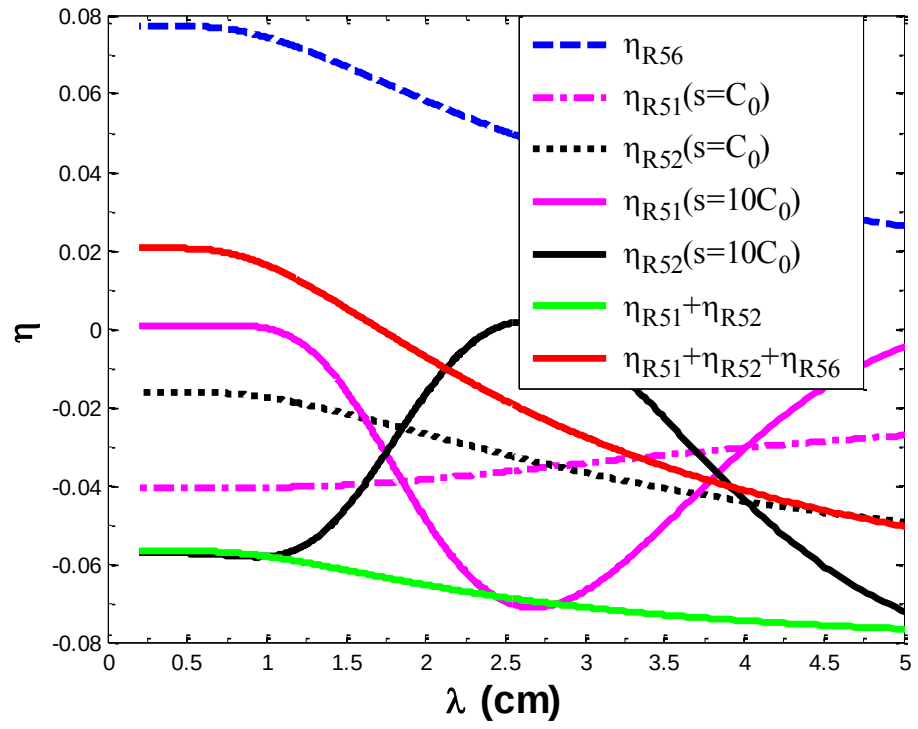

Fig. 2. Slip factors for $I_0 = 10\ \mu$A at $s=C_0$ and $s=10\ C_0$.

Fig. 1 and Fig. 2 demonstrate that, for a beam in an isochronous ring, because of smallness of $\eta_{R56}$, the component of the space-charge modified slip factor $\eta_{R51}+\eta_{R52}$ contributed from the elements of $R_{51}$ and $R_{52}$ plays an important role in the longitudinal beam dynamics and cannot be neglected. The situation is different from a storage ring working far from transition. Note in Fig. 1 and Fig. 2, the total slip factor $\eta_{R51}+\eta_{R52}+\eta_{R56}$ can be negative for some given perturbations wavelengths and beam parameters.

### *4.2. Exponential suppression factor*

We can define the exponential function of the integrand in Eq. (41) as exponential suppression factor (E.S.F.)

$$E.S.F. = e^{-(k\sigma_x/\nu_{inc})^2[1-\cos(\nu_{inc}\chi/R)]^2} \times e^{-\frac{1}{2}[\frac{k\sigma_E}{\beta^2 E}(1/\gamma_{t,inc}{}^2-1/\gamma^2)\chi]^2}. \quad (51)$$

The first exponential function in Eq. (51) originates from the smooth approximation of $\Delta R_{51}(s', s) = R_{51}(s) - R_{51}(s')$, $\Delta R_{52}(s', s) = R_{52}(s) - R_{52}(s')$ and the emitttance $\varepsilon_{x,0} = \sigma_x^2 \nu_{x,inc}/R$. While the second exponential function in Eq. (51) originates only from the RMS energy spread and $\Delta R_{56}(s', s) = R_{56}(s) - R_{56}(s')$ without the contributions of $R_{51}$ and $R_{52}$. The *E.S.F.* is a measure of Landau damping effects for the microwave instability of SIR beam.

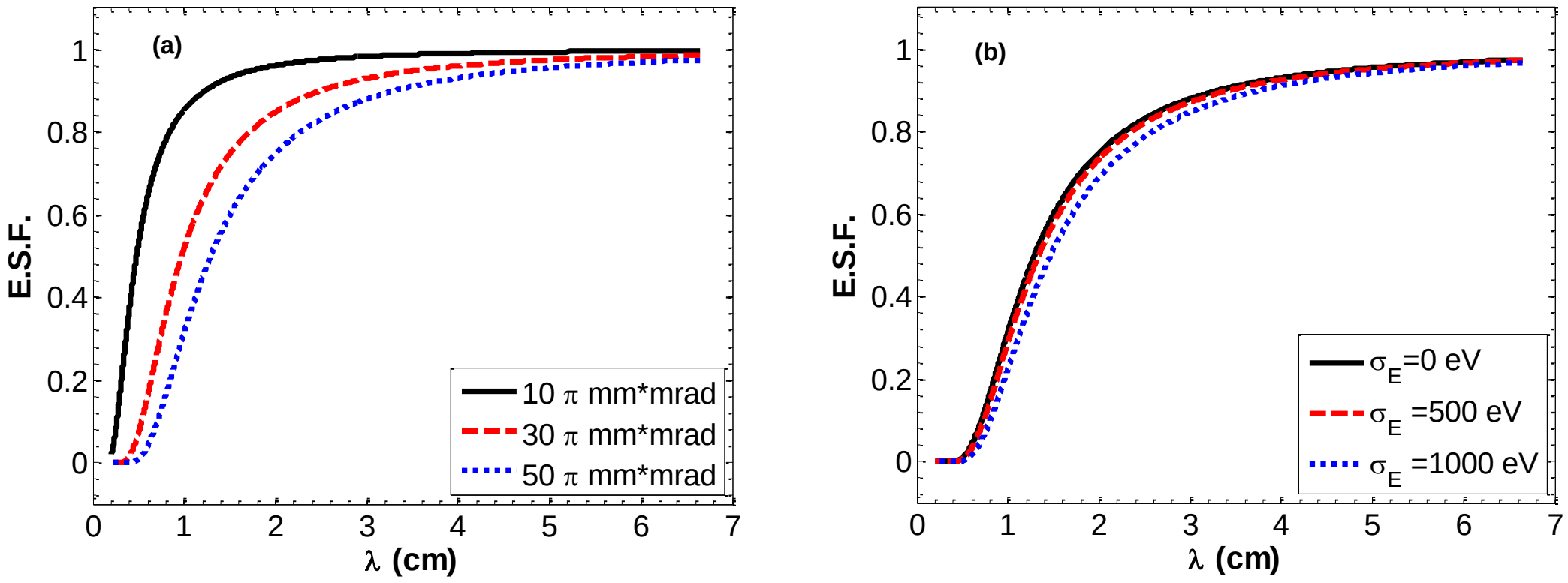

Fig. 3. The *E.S.F.* at $\chi = C_0$ for a 1.0 $\mu$A, 19.9 keV SIR beam. (a) $\sigma_E = 0$, and variable emittance. (b) $\varepsilon_{x,0}$= 50 $\pi$ mm mrad, and variable $\sigma_E$.

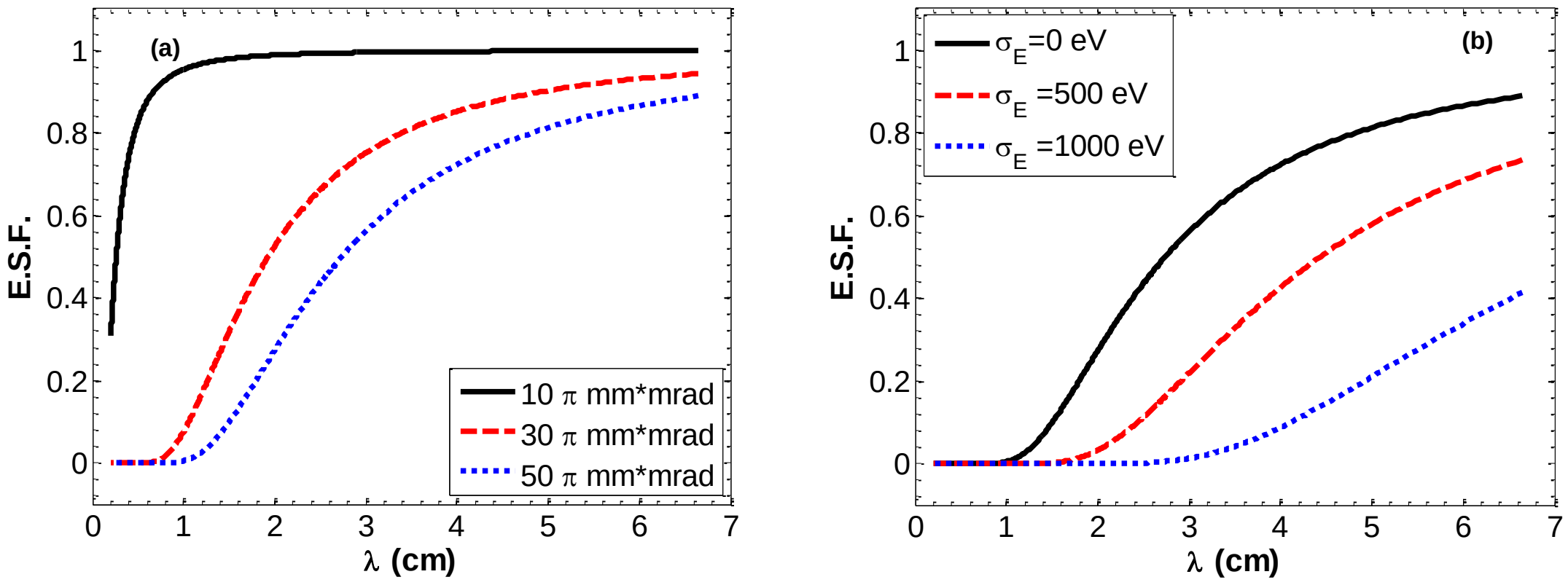

Fig. 4. The *E.S.F.* at $\chi = 10\ C_0$ for a 1.0 $\mu$A, 19.9 *keV* SIR beam. (a) $\sigma_E = 0$, and variable emittance. (b) $\varepsilon_{x,0}$= 50 $\pi$ mm mrad, and variable $\sigma_E$.

For a SIR beam with the current of 1.0 $\mu$A, mean kinetic energy of 19.9 keV, Fig. 3 (a) shows the calculated *E.S.F.* at $\chi = C_0$ with $\sigma_E = 0$ and variable emittance. Fig. 3 (b) shows the calculated *E.S.F.* at $\chi = C_0$ with $\varepsilon_{x,0}$=50 $\pi$ mm mrad and variable $\sigma_E$. Note for a beam without uncorrelated energy spread ($\sigma_E = 0$ eV), the *E.S.F.* in the short-wavelength limits is still small due to the finite beam emittance effect. Since the *E.S.F* is related to $exp[-A(k\sigma_x)^2]=exp[-A(2\pi\sigma_x/\lambda)^2]$ and $exp[-B(k\sigma_E)^2]=exp[-B(2\pi\sigma_E/\lambda)^2]$, where $A$ and $B$ are coefficients which are independent of $\sigma_x$, $\sigma_E$ and $\lambda$, then the Landau damping effects are more effective for a beam with large uncorrelated RMS energy spread and emittance at the shortest perturbation wavelengths. Fig. 4(a) and Fig. 4(b) show the calculated *E.S.F.* at $\chi = 10\ C_0$. Comparison between Fig. 3(b) and Fig. 4(b) indicates the *E.S.F.* decreases significantly with larger relative path length difference $\chi$ due to the finite uncorrelated energy spread effect.

The radial-longitudinal coupling matrix elements $R_{51}$ and $R_{52}$ may affect the microwave instability in an isochronous ring in two ways. (a) Eqs. (47)-(50) show $R_{51}$ and $R_{52}$ may modify the coherent space-charge modified slip factor for a beam with coherent energy deviations and associated radial centroid offsets. (b) Eq. (51) shows, if a coasting beam has finite energy spread and emittance, the incoherent motions of charged particles under the effects of matrix elements $R_{51}$, $R_{52}$ and $R_{56}$ may produce a finite spread in the longitudinal motion spectrum around the revolution frequency. The revolution frequency spread can help to smear out the longitudinal charge density modulations and suppress the microwave instability growth rates, especially for the short-wavelength perturbations. This is the origin of the Landau damping effects in the isochronous regime. Since the matrix elements $R_{51}$, $R_{52}$ and $R_{56}$ may affect the beam instability in such a complicated way, it is difficult to predict how the instability growth rates will change if one of these elements is increased or decreased.

### *4.3. Relations between the 1D growth rates formula and 2D dispersion relation*

In the 2D dispersion relation of Eq. (41), if we neglect the *E.S.F.* and the sinusoidal term in the square bracket of the integrand which originates from the coupling matrix elements $R_{51}(s)$ and $R_{52}(s)$, the 2D dispersion relation is reduced to

$$1 = -\frac{ie\Lambda_0}{\beta\gamma m_{H_2+}cC_0} kZ_{0,sc}^{\parallel}(k)\int_0^{\infty} d\chi e^{-\mu\chi}\left(\frac{1}{\gamma_{t,coh}^2} - \frac{1}{\gamma^2}\right)\chi. \qquad (52)$$

With Eq. (42) and the equality of

$$\int_0^{\infty} dse^{-\mu s}s = \frac{1}{\mu^2}, \qquad (53)$$

the simplified 2D dispersion relation Eq. (52) can be reduced further as

$$\tau^{-1}(k) = \mu\beta c = \omega_0 \sqrt{-i\frac{\eta_{coh} e I_0 k R Z_{0,sc}^{\parallel}}{2\pi\beta^2 E}}. \qquad (54)$$

Eq. (54) is just the Eq. (1) for a 1D mono-energetic and laminar beam.

Though the model and the 1D dispersion relation in Ref. [5] can predict the fastest-growing wavelength, the predicted growth rates are not proportional to the unperturbed beam intensity $I_0$. In Ref. [5], the longitudinal line density is $N(z) = N_k cos(kz)$, and the radial coherent space charge field factor $\alpha$ calculated in Eq. (12) of Ref. [5] is proportional to $N(z)$. In Eq. (23) of Ref. [5], the constant parameter is proportional to the unperturbed line density $N_0$ which is from Eq. (18) for the longitudinal beam dynamics. Considering Eq. (24) of the same paper, since the instability growth rates $\omega_i$ predicted by Eq. (23) are proportional to $[N_0 N_k(z)]^{1/2}$ instead of $N_0$ or $I_0$, then the predicted instability growth rates of this model and theory violate the scaling law with respect to the unperturbed beam intensity $I_0$ observed in our experiments and simulations. In reality, the longitudinal line density should be $N(z)=N_0+N_k cos(kz)$, the above discrepancy results from the missing of $N_0$ in the model in calculation of the radial space charge field factor $\alpha$. Note the parameter $\chi_2(k)$ in Eq. (17) and Eq. (23) of Ref. [5] has a similar behavior to $1 - kr_0 \cdot K_1(kr_0)$ plotted in Fig. 5 of Ref. [4], which peaks at wavelength $\lambda = 0$ and decreases monotonically with $\lambda$. If $N_0$ were included in this model, this model would be similar to the one in Ref. [3, 4]. It can neither explain the suppression of the short-wavelength perturbations nor predict the fastest-growing wavelengths properly.

## 5. Simulation study of the microwave instability in SIR

### *5.1. Simulated growth rates of microwave instability*

In this section, we will study the simulated spectral evolutions of the microwave instability using the Fast Fourier Transform (FFT) technique and compare with the theoretical calculations. Studies of the long-term microwave instability are carried out by running CYCO up to 100 turns for a macro-particle bunch to mimic a real $H_2^+$ bunch in SIR. The bunch has an initial beam intensity $I_0 = 1.0\ \mu$A, mono-energetic kinetic energy $E_{k0} = 19.9$ keV, radial and vertical emittance $\varepsilon_{x,0} = \varepsilon_{y,0} = 50\ \pi$ mm mrad. The initial distributions are uniform in both the 4-D transverse phase space ($x, x', y, y'$) and the longitudinal charge density. A total of 300000 macro-particles are used in the simulation. Considering both the curvature effects on the space charge fields when a long bunch enters the bending magnets, and the edge field effects of a short bunch, a bunch length of $\tau_b = 300\ ns$ ($L_b \approx 40$ cm) is selected in the simulation. Due to the strong nonlinear edge effects in the bunch head and tail, only the beam profiles of the central part of the bunch with longitudinal coordinates -10 cm ≤ $z$ ≤ 10 cm are analyzed by FFT. At each turn, the density profiles of the coasting bunch with coordinates of -10 cm ≤ $z$ ≤ 10 cm are sliced into 512 small bins along $z$, the number of macro-particles in each bin was counted, and the 512-point FFT analysis is performed with respect to $z$. The microwave instability of SIR beam is a phenomenon of line charge density perturbations with typical wavelengths of only several centimeters. The full chamber height is about 4.8 cm, which gives the approximate upper limit of the perturbation wavelengths in the simulation study. According to the Nyquist - Shannon Sampling Theorem, the shortest wavelength which can be analyzed by the 512-point FFT is 2×20 cm/512 ≈ 0.078 cm. Since the beam diameter is about 1.0 cm, the simulation results for the shorter wavelengths of $\lambda$=1.0 cm, 0.5 cm and 0.25 cm may give us some insights on the instabilities of short wavelengths comparable to or less than the transverse beam size. A series of mode number of 4, 5, 7, 10, 20, 40 and 80 are selected for the 20-cm-long beam profiles, which give the corresponding line charge density perturbation wavelengths of $\lambda$=5 cm, 4 cm, 2.857 cm, 2 cm, 1 cm, 0.5 cm and 0.25 cm. The growth rates of these wavelengths are fitted numerically and studied in the analysis. In order to minimize the effects of randomness in the initial beam micro-distribution on the simulation results, for each setting of beam parameters, the code CYCO is run 5 times for 5 different initial micro-distributions, and the average growth rates of the 5 runs for each given perturbation wavelength are used as the simulated growth rates in the analysis work.

Fig. 5-Fig. 7 show the simulated beam profiles and line density spectrum at turn 0, turn 60, and turn 100 for a single run of CYCO, respectively. In each of these figures, the left graph displays the top view of the beam distributions (blue dots) superimposed by the number of macro-particles per bin (red curve); the right graph displays the spectrum of the line charge density analyzed by FFT. We can see the line density

modulation amplitudes increase with turn numbers, and the peaks of the line density spectrum shift to the frequencies around $1/\lambda \approx 0.5\ \text{cm}^{-1}$ or the wavelength $\lambda \approx 2.0$ cm at large turn number.

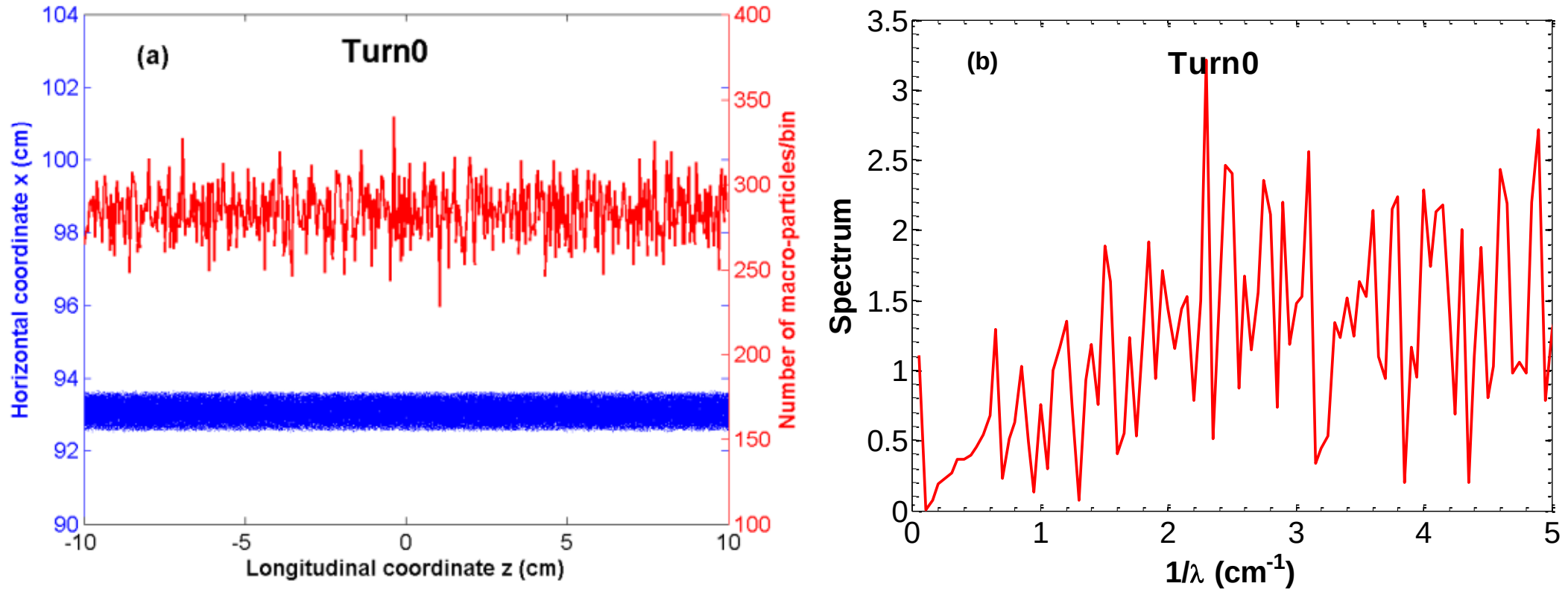

Fig. 5. (a) Beam profiles and (b) line density spectrum at turn 0.

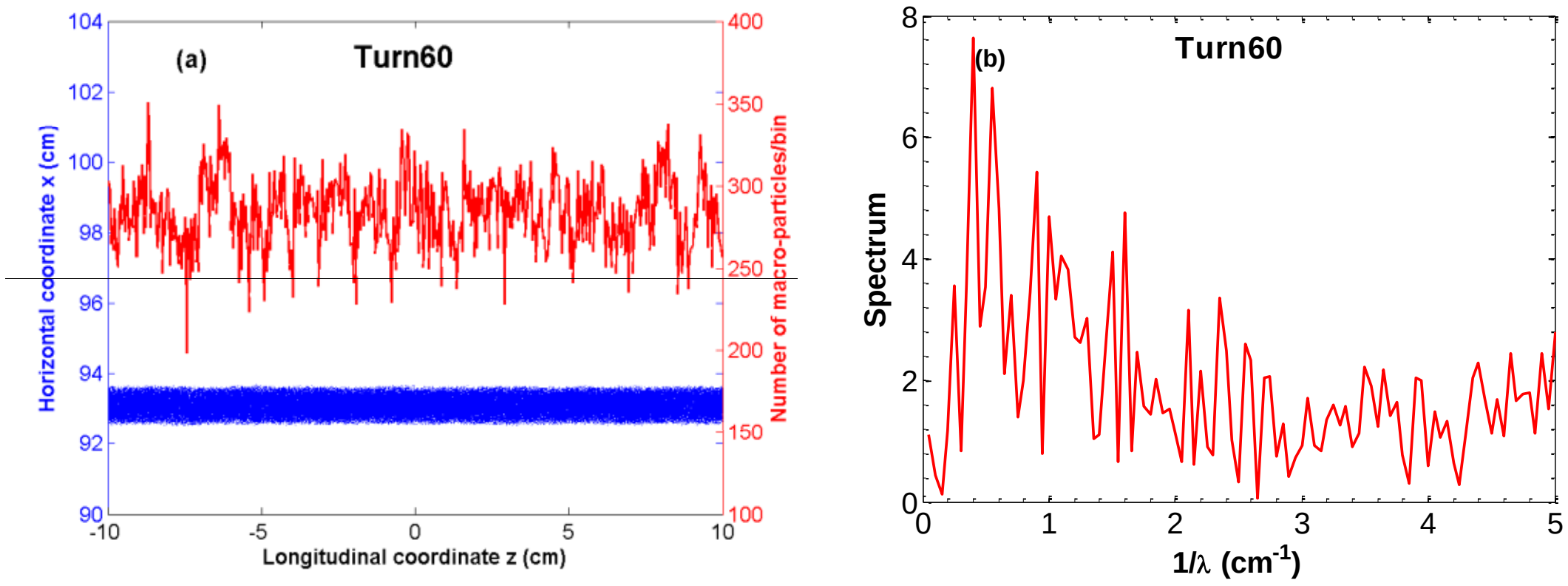

Fig. 6. (a) Beam profiles and (b) line density spectrum at turn 60.

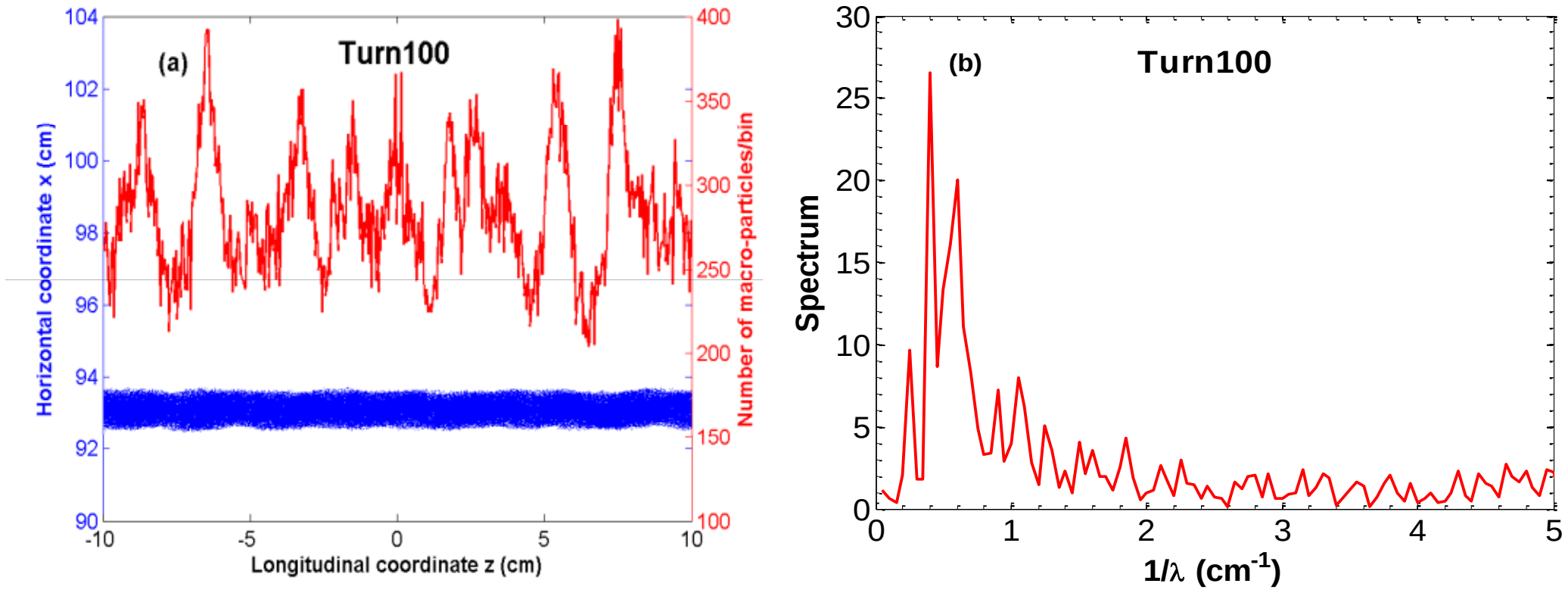

Fig. 7. (a) Beam profiles and (b) line density spectrum at turn 100.

Fig. 8 shows the FFT analysis results of the temporal evolutions of the normalized line charge densities $\hat{\Lambda}_k / \Lambda_0$ for the seven chosen line charge density perturbation wavelengths $\lambda$ up to turn 100 for a single run of CYCO. We can see there are some oscillations superimposed on the exponential growth curves.

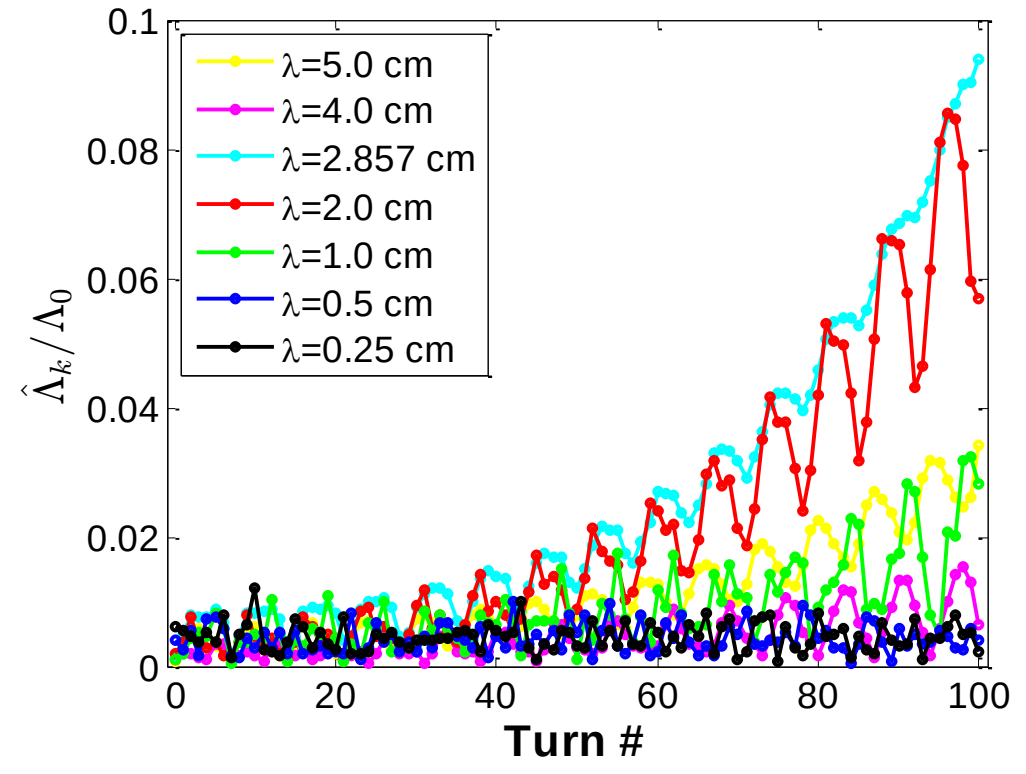

Fig. 8. Evolutions of harmonic amplitudes of the normalized line charge densities.

Fig. 9 shows the temporal evolutions of the normalized line charge densities spectrum of given wavelengths and the fitting results using proper fitting functions.

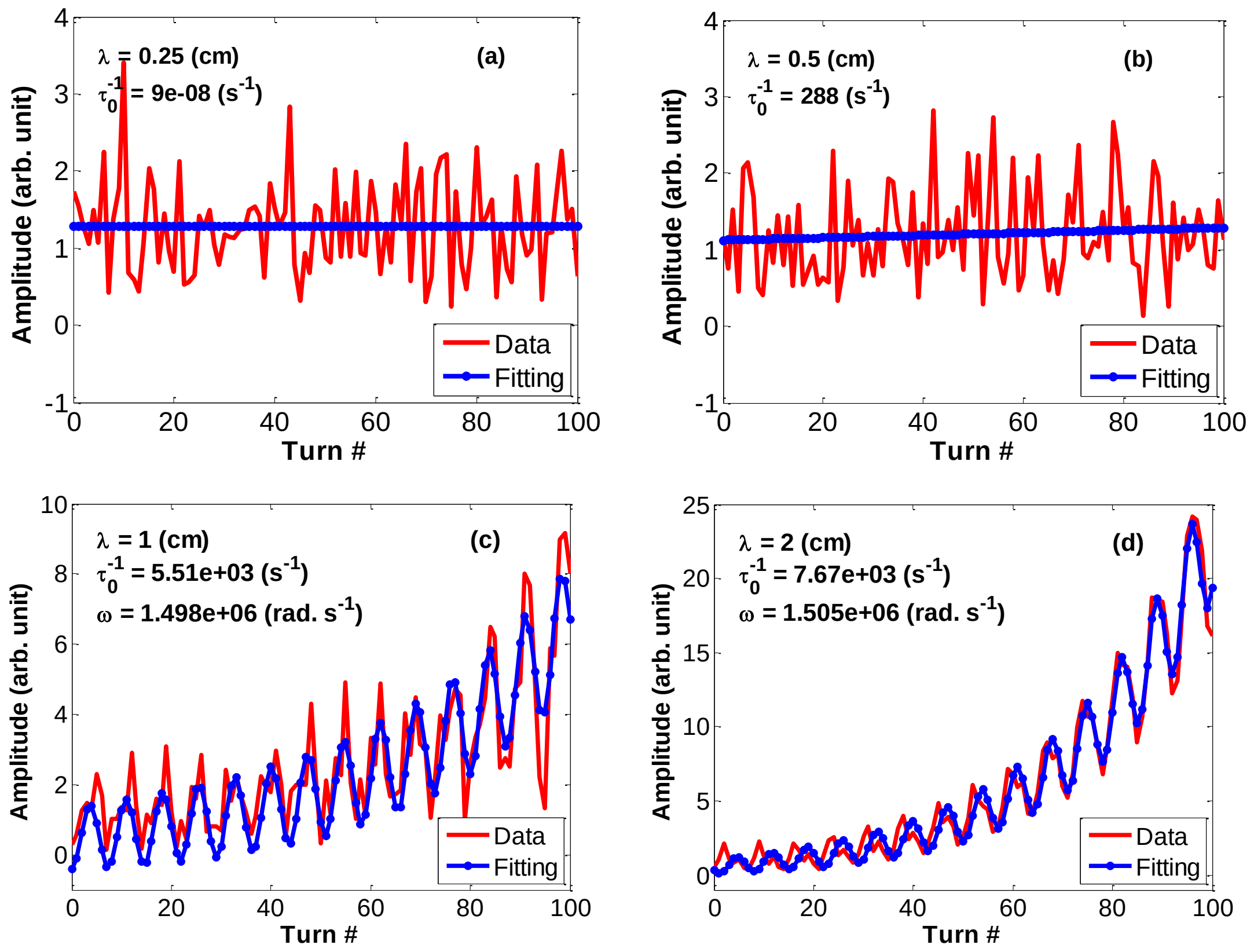

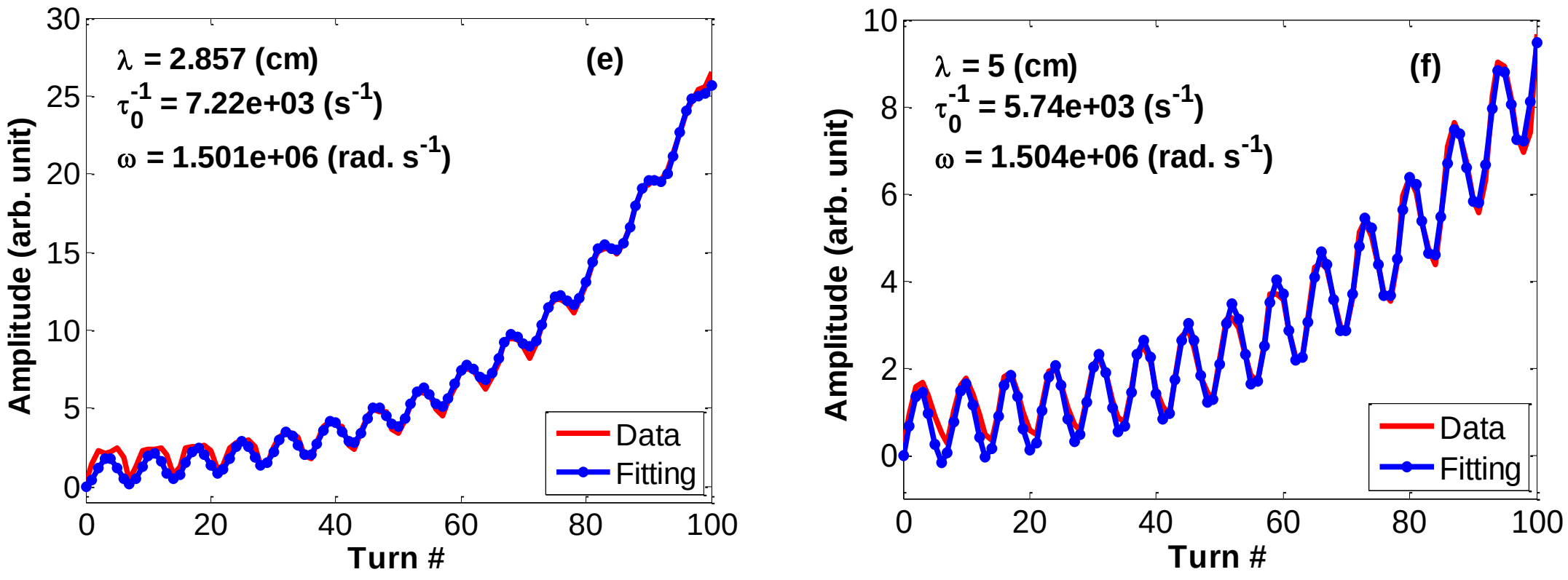

Fig. 9. Curve fitting results for the growth rates of the normalized line charge densities for a single run of CYCO. (a) $\lambda$ = 0.25 cm; (b) $\lambda$ = 0.5 cm; (c) $\lambda$ = 1.0 cm; (d) $\lambda$ = 2.0 cm; (e) $\lambda$ = 2.857 cm; (f) $\lambda$ = 5.0 cm.

For the cases of $\lambda$ = 0.25 cm and $\lambda$ = 0.5 cm, since the oscillations are irregular, we choose the fitting function as

$$|\Lambda_1(t)| \approx \hat{\Lambda} e^{\frac{t}{\tau_0}}. \tag{55}$$

For the cases of $\lambda$ =1 cm, 2 cm, 2.857 cm, 4 cm, and 5 cm, since there are obvious sinusoidal oscillations in line charge densities superimposed on exponential growths, we choose the fitting function as

$$|\Lambda_1(t)| \approx \hat{\Lambda} e^{\frac{t}{\tau_0}} + P e^{Qt/T_0} \cos(\omega t + \Phi), \tag{56}$$

where $\hat{\Lambda}$, $P$, $Q$, $\omega$, $\Phi$, and $\tau_0$ are the fit coefficients, $T_0$ is the revolution period of $H_2^+$ ion, $t = N_t T_0$, $N_t$ is the turn number, $\tau_0^{-1}$ is just the long-term instability growth rate. Note for beam energy of 19.9 keV, the nominal angular betatron frequency is $\omega_\beta$ = 1.499 × 10$^6$ radian/second. The fitting results show the oscillations in the curves for $\lambda$ =1 cm, 2 cm, 2.857 cm, and 5 cm are just the betatron oscillations, they are the dipole modes in the longitudinal structure of the beam due to dipole moment of the centroid offsets [6].

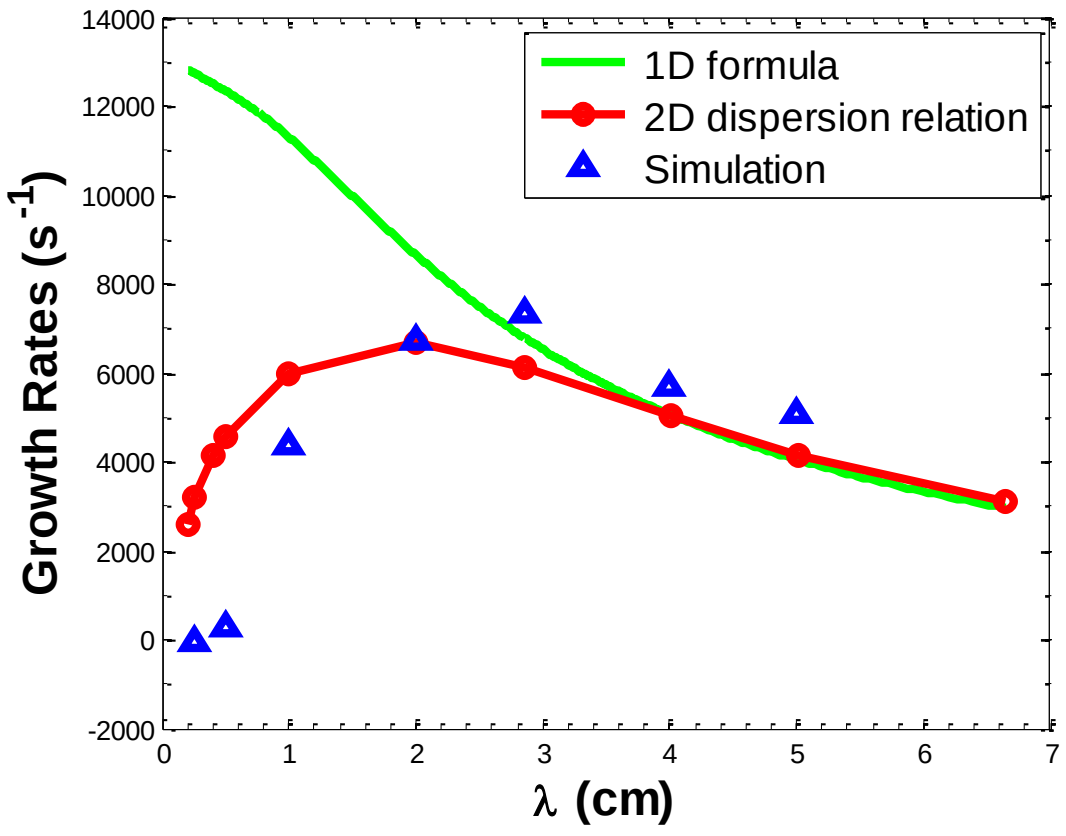

Fig. 10. Comparison of the instability growth rates between theory and simulations for 5 runs of CYCO.

Fig. 10 shows the comparison of the microwave instability growth rates between the theoretical values and the average simulation results for 5 runs of CYCO. Note that the theoretical values are predicted by the 1D formula of Eq. (1) with the slip factor expressed in Eq. (4) plus $\eta_0$, and the 2D dispersion relation of Eq. (41) with the space-charge modified tunes and transition gammas expressed in Eqs. (13) - (16). For both the 1D and 2D formalisms, the LSC impedances are calculated by Eq. (3), and the beam radii $r_0$ are calculated from the solution to the algebraic matched-beam envelope Eqs. (4.93) of Ref. [7]. Note that in Fig. 10, the 1D formalism used in Refs. [3, 4] and the 2D dispersion relation have similar performance in prediction of the growth rates of the long-wavelength perturbations ($\lambda \geq$ 4 cm), which are all consistent with the simulation results. For $\lambda$<2 cm, the 1D formalism significantly overestimates the instability growth rates as $\lambda \to 0$ and cannot predict the fastest-growing wavelength ($\lambda \approx$ 2 *cm*) correctly, because Eq. (1) neglects the Landau damping effects of finite emittance and energy spread; while the 2D dispersion relation, with the Landau damping effects taken into account, has a much better performance than the conventional 1D formula in prediction of instability growth rates in the short-wavelength limits ($\lambda$<2 cm) and the fastest-growing wavelength, though there still exist some bigger discrepancies between the simulations and theory for very short wavelength $\lambda$<1 cm. Therefore we can see the Landau damping is a necessary mechanism to explain the low instability growth rates of the short-wavelength perturbations ($\lambda$ is less than or comparable to $r_0$), which cannot be explained by the conventional 1D formalisms. Only at larger wavelengths ($\lambda >> r_0$) will the 1D and 2D dispersion relations have the similar performance.

### *5.2. Growth rates of instability with variable beam intensities*

In order to study the dependence of microwave instability growth rates on initial beam intensities, simulations using CYCO are carried out for SIR beams with $E_{k0}$ = 19.9 keV, $\sigma_E$= 0 eV, $\tau_b$ = 300 *ns* ($L_b \approx$ 40 cm), $\varepsilon_{x,0} = \varepsilon_{y,0}$ = 50 $\pi$ mm mrad, $I_0$ = 0.1, 0.3, 0.5, 5.0, and 20 $\mu$A, respectively. The initial distributions are uniform in both the 4-D transverse phase space ($x$, $x'$, $y$, $y'$) and the longitudinal charge density. The simulation for each intensity level is performed 5 times using 5 different initial micro-distributions, and the average simulated growth rates of the selected perturbation wavelengths of the 5 runs are used in analysis. For $I_0$ > 20 $\mu$A, due to fast development of beam instability, the beam dynamics may enter the nonlinear regime only after several turns of coasting, this makes the fitting work difficult and inaccurate, therefore the simulation results for the intensities of $I_0$ > 20 $\mu$A are not available in this paper.

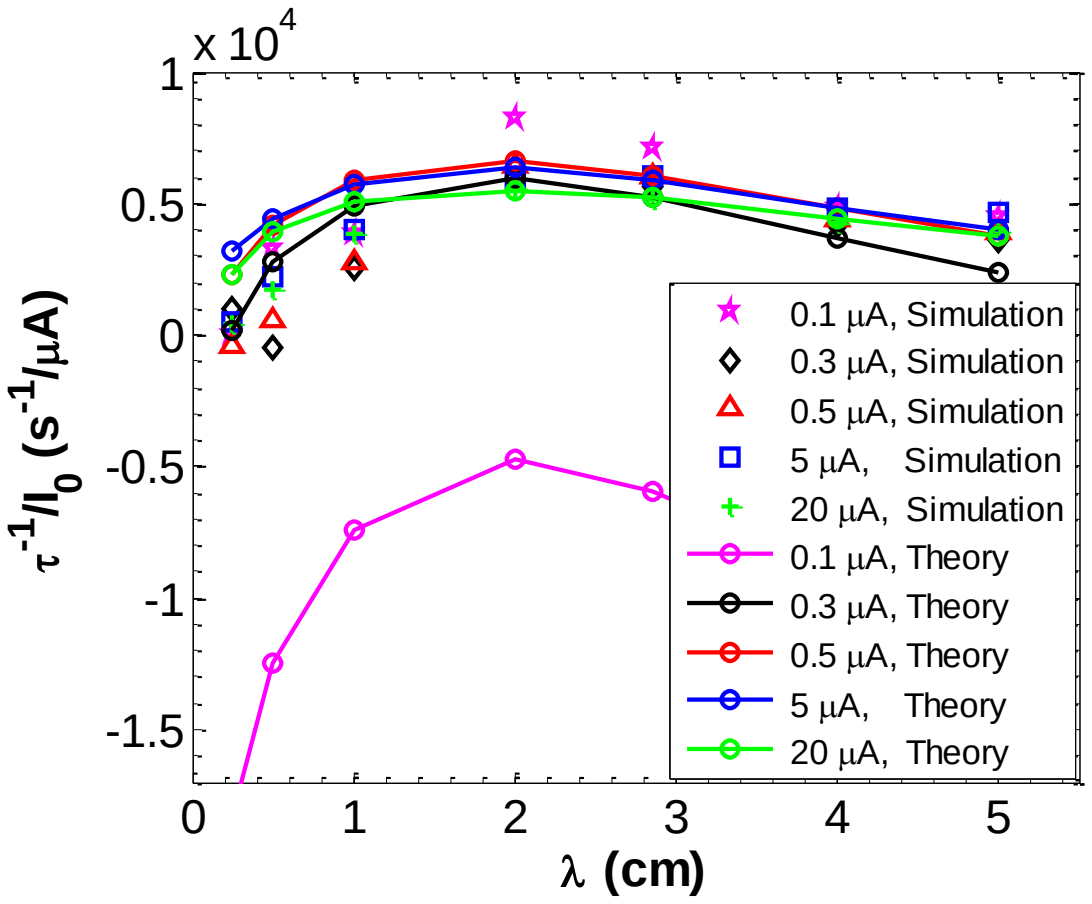

Fig. 11. Comparisons between the simulated and theoretical normalized instability growth rates for different beam intensities.

Fig. 11 shows the comparisons between the simulated and theoretical normalized instability growth rates for beam intensities ranging from 0.1 $\mu$A to 20 $\mu$A. We can see, except for $I_0$ = 0.1 $\mu$A, the theoretical normalized growth rate curves roughly overlap each other within a region. The theory and simulations are roughly consistent to each other for $\lambda$ > 2 cm and 0.3 $\mu$A $\leq I_0 \leq$ 20 $\mu$A. For $\lambda$ < 2 cm, the discrepancies between the simulation and theory become bigger.

### *5.3. Growth rates of instability with variable beam emittance*

In order to study the dependence of microwave instability growth rates on initial beam emittance, simulations using CYCO are carried out for SIR beams with $E_{k0}$ = 19.9 keV, $\sigma_E$= 0 eV, $\tau_b$ = 300 *ns* ($L_b \approx$ 40 cm), $I_0$ =1.0 $\mu$A, $\varepsilon_{x,0}$ = $\varepsilon_{y,0}$ = 30, 50 and 100 $\pi$ mm mrad, respectively. The code CYCO is run up to 100 turns and the growth rates are fitted by proper functions. For each initial emittace, the average growth rates of 5 runs with different initial micro-distributions are used in the analysis. Fig. 12 shows the comparisons of growth rates between theory and simulations. We can see for $\lambda$ > 1.0 cm, the simulated and theoretical instability growth rates are consistent with each other, the larger emittance may help to suppress the instability growth rates. While for $\lambda$ < 1.0 cm, the discrepancies between the simulation and theory become bigger.

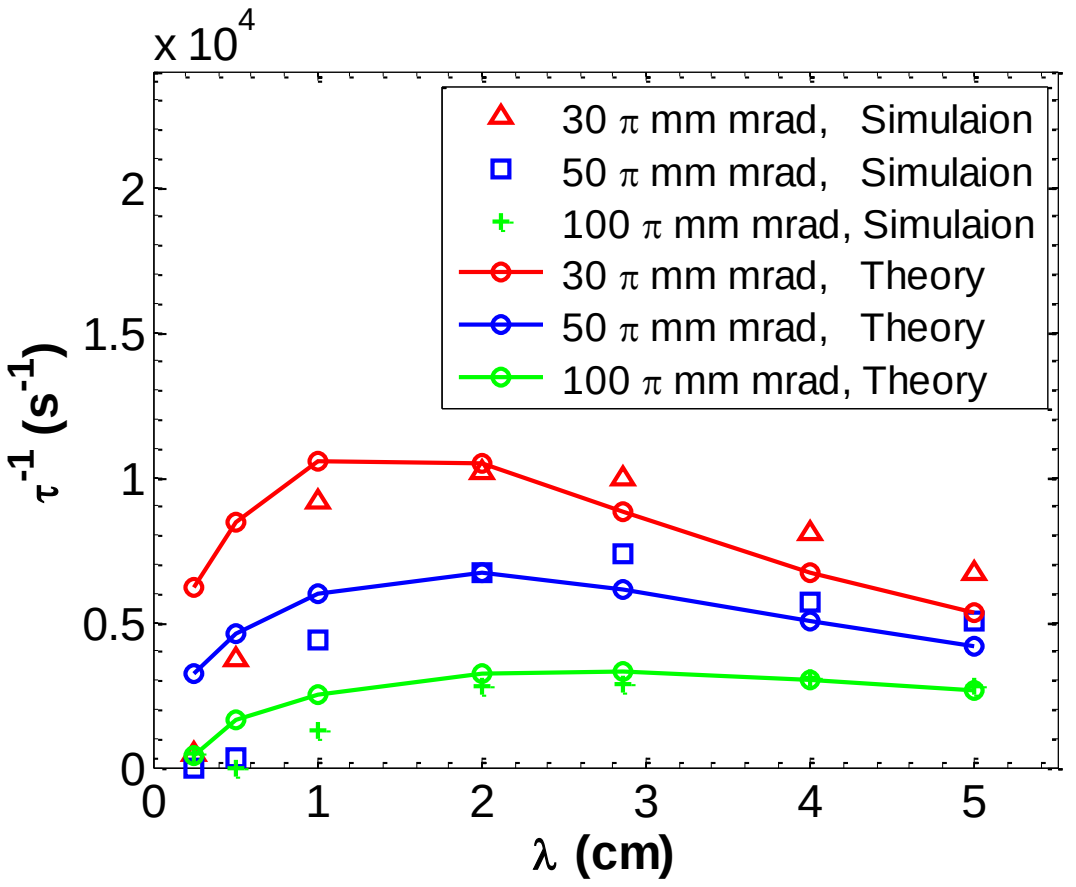

Fig. 12. Comparisons of microwave instability growth rates between theory and simulations for variable initial emittance.

### *5.4. Growth rates of instability with variable beam energy spread*

Fig. 13 shows the comparisons of growth rates between theory and simulations for SIR beams with $E_{k0}$ = 19.9 keV, $\tau_b$ = 300 *ns* ($L_b \approx$ 40 cm), $I_0$ =1.0 $\mu$A, $\varepsilon_{x,0}$ = $\varepsilon_{y,0}$ = 50 mm mrad, $\sigma_E$ = 0, 50, and 75 eV, respectively. The code CYCO was run up to 100 turns and the growth rates are fitted by proper functions. For each initial RMS energy spread $\sigma_E$, the average growth rates of 5 runs with different initial micro-distributions are used in the analysis. We can see for $\lambda$ >2.0 cm, the simulated and theoretical instability growth rates are consistent with each other, the larger energy spread may help to suppress the instability growth rates. While for $\lambda$ < 2.0 cm, the discrepancies between the simulation and theory become bigger.

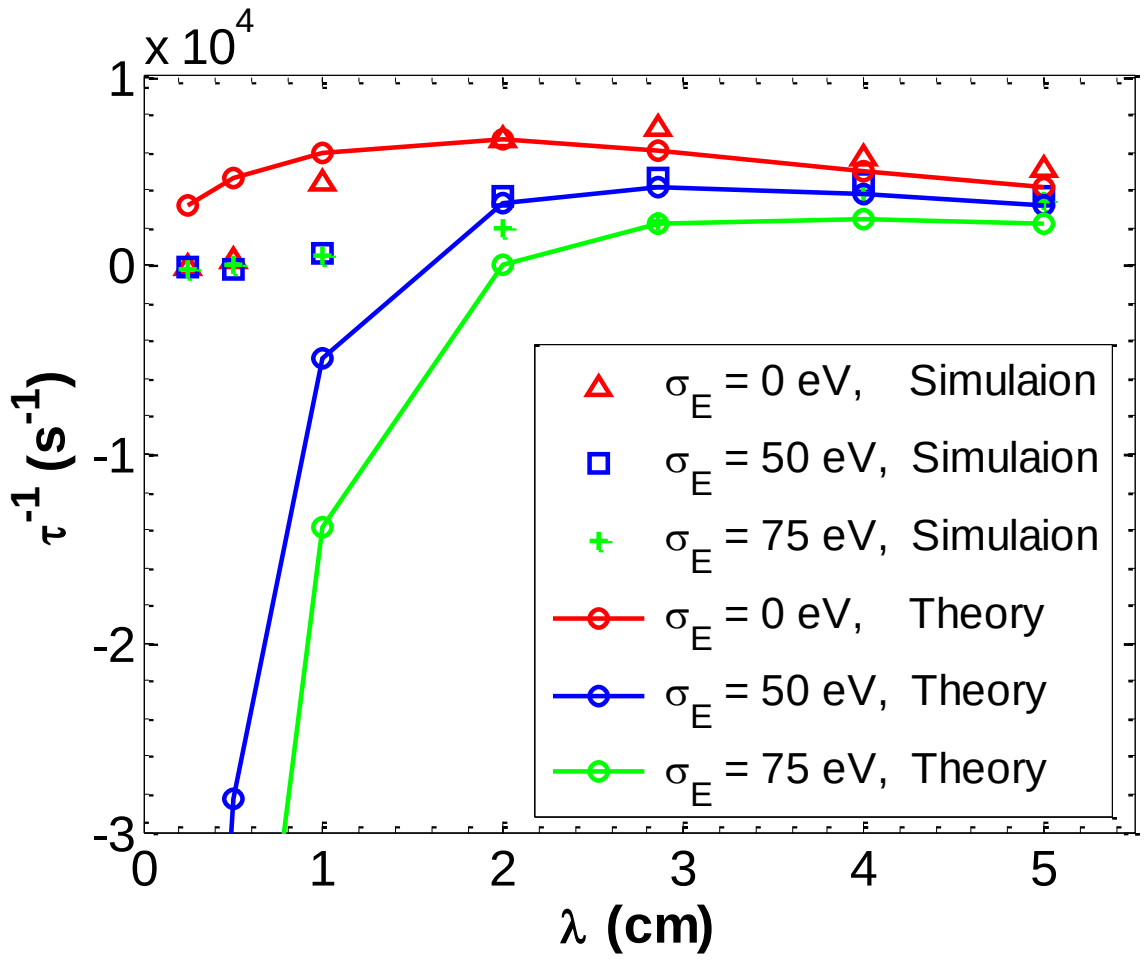

Fig. 13. Comparisons of microwave instability growth rates between theory and simulations for variable uncorrelated RMS energy spread.

*5.5. Possible reasons for the discrepancies between simulations and theory in the short-wavelength limits*

Figs. (10)-(13) show there exist bigger discrepancies between the theoretical and simulated instability growth rates in the short-wavelength limits (especially for $\lambda \leq 1.0$ cm), they may be caused by one or some of the following reasons:

(a) Deviation from the beam model

The 2D dispersion relations Eqs. (30)(41) are derived from the unperturbed Gaussian beam distribution described in Eqs. (17) and (18), which can be rewritten as product of three Gaussian distribution functions

$$f_0 = \frac{n_b}{(2\pi)^{\frac{3}{2}}\varepsilon_{x,0}\sigma_\delta}\exp\left(-\frac{x_0^{\ 2}}{2\varepsilon_{x,0}\hat{\beta}_0}\right)\exp\left[-\frac{(x_0')^2}{2\frac{\varepsilon_{x,0}}{\hat{\beta}_0}}\right]\exp\left[-\frac{(\delta+\hat{u}z_0)^2}{2\sigma_\delta^2}\right], \tag{57}$$

The model assumes the transverse phase space ($x_0$, $x_0'$) is centered at ($<x_0>=0$, $<x_0'>=0$). For storage rings, the assumption of the linear chirp factor $\hat{u}=0$, the compression factor $C(s)=1$ are also used in the derivations. Therefore the coherent fractional momentum deviation $<\delta_0>=<\delta+\hat{u}z_0>=<\delta>=0,$ and there is no correlation between the transverse and longitudinal distributions. The initial unperturbed distribution function described in Eq. (57) is just product of three normal distribution functions with zero-mean. While as the beam coasts in the ring, there will be local centroid offset $<x_0>$ induced by the coherent fractional momentum deviation $<\delta_0>$ due to dispersion function $D$:

$$<x_0>=\mathrm{D}<\delta_0>. \tag{58}$$

In addition, when sinusoidal centroid wiggling takes place due to space charge force, the correlated fractional moment deviation $\hat{u}z_0$ should be replaced by a sinusoidal function of $z_0$, then the compression factor $C(s)\neq1$ and will be dependent on $s$ or $t$. The non-zero $<x_0>$, $<\delta_0>$ and non-constant, $s$-dependent compression factor $C(s)$ will shift the centers of bean distributions in the longitudinal and transverse phase space, produce a transverse-longitudinal correlation in distribution function. Consequently, the 2D dispersion relation Eqs. (30)(41) will be modified, the amplitude of perturbed harmonic line density $|n_{1,k}(z,t)|=|g_k(t)|$ described in Eq. (34) should be replaced by $|n_{1,k}(z,t)|=|C(t)g_k(t)|$ too. This may cause the bigger discrepancies between the theoretical and simulated instability growth rates in the short-wavelength limits.

(b) Curvature effects

The LSC impedance, space-charge modified betatron tunes and transition gammas are all derived for an infinite long, straight beam model, while the SIR consists of four 90-degree bending magnets which account for about 43% of the ring circumference. When the beam enters these bends, the curvature effects on the longitudinal and transverse beam dynamics are not taken into account in the theoretical analysis.

(c) LSC fields of the dipole mode

The centroid wiggling may induce the LSC fields of the dipole mode which are neglected in the theoretical analysis.

(d) Spectral leakages

Because the line charge density perturbations around the fastest-growing wavelength ($\lambda \approx 2.0$ cm) have larger amplitudes comparing to the modes with smaller growth rates, and the FFT analysis is applied to a bunch with finite length using rectangular window, these fastest-growing modes may inevitably create the new frequency components (false spectrum) spreading in the whole frequency domain, namely, the so-called spectral leakages. The leaked spectra from the faster-growing modes may mix with and mask the real spectra of the slower-growing modes, therefore lower the resolutions of the FFT analysis results.

(e) Initial distribution

In Figs. (12)(13), because the beams with uniform longitudinal charge density are used in the simulations, their residual line charge modulation amplitudes are vanishingly small (theoretically speaking, they should be 0 in ideal conditions). When the growth rates in short-wavelength limits are negative due to larger beam emittance and energy spread, they can hardly be detected since the initial density modulation amplitudes have reached minimum already.

In summary, the bigger discrepancies between the theoretical and simulated instability growth rates in the short-wavelength limits may be caused by various reasons, due to complexity of the problem, for the time being, further discussions are not available in this paper.

## 6. Long-term evolutions of energy spread

For a coasting bunched beam, the evolution of the energy spread induced by the space charge field plays an important role in both the linear and nonlinear regime of beam instability. A large energy spread may help to suppress the microwave instability in the linear beam dynamics, it is also one of the important measures of the asymptotic bunch behavior in the nonlinear beam dynamics. When a high intensity uniform long $H_2^+$ bunch with a finite length is injected into the SIR, the nonlinear space charge forces in the beam head and tail are strong and may deform the beam shape. In addition, the bunch may break up into many small clusters longitudinally only after several turns of coasting due to the microwave instability [1, 2]. Therefore the beam dynamics is highly nonlinear in these cases. In this section, we mainly discuss the energy spread evolutions in the nonlinear dynamics by simulations and experimental methods.

SIR lab has designed a compact electrostatic retarding field energy analyzer [19], which can scan across the beam radially to measure the energy spread of a bunch at various chosen number of turns after injection. The analyzer is roughly a cube with a 1 mm (width) × 14 mm (height) entrance slit located in the middle of the entrance plate. It consists of a rectangular housing tube, a rectangular high voltage retarding tube, a fine retarding mesh (1000 lines per inch, 50% transmission rates), a secondary electron suppressor, and a current collector. The analyzer is installed below the median ring plane in the extraction box. The entrance plate is tilted at an angle with respect to the vertical plane to align the analyzer axis parallel to the deflected beam. A pair of high-voltage pulsed electrostatic deflectors is used to kick the beam down to the energy analyzer when the energy spread measurements are performed. Usually, the energy spread is measured at three radial positions: one is at the location of the peak beam current, and the other two are close to the beam core edges on each side.

A $H_2^+$ ion bunch with length 600 mm, peak current 8.0 $\mu$A, kinetic energy 10.3 keV and emittance 30 $\pi$ mm mrad is used in the energy spread measurements. A mono-energetic macro-particle bunch of the above initial parameters with a uniform distribution in both the longitudinal line charge density and the 4D transverse phase space is also used in the simulation study by the code CYCO. In the analysis of simulation results, the radial beam is cut into several 1-mm-wide small bins, and the number of macro-particles, mean kinetic energy and RMS energy spread in each bin are calculated and compared with the experimental values. Fig. 14 shows the simulated and experimental radial slice beam densities. Fig. 15 shows the simulated top views and RMS slice energy spread at turn 4 and turn 30, respectively. Fig. 16 shows the simulated RMS slice energy spread up to turn 8. Fig. 17 shows the comparison of RMS slice energy spread between simulations and experiments. Note that in this paper the slice energy spread and slice density denote all the slices are cut parallel to the longitudinal $z$ direction instead of the radial direction, which is conventionally used in free-electron lasers (FELs). Considering the charged long bunch is a chaotic system, a small difference in the initial beam distribution may cause a huge beam profile deviation at large turn numbers, we can see that the simulated radial beam density profiles and RMS slice energy spread match the experimental values within an acceptable range.

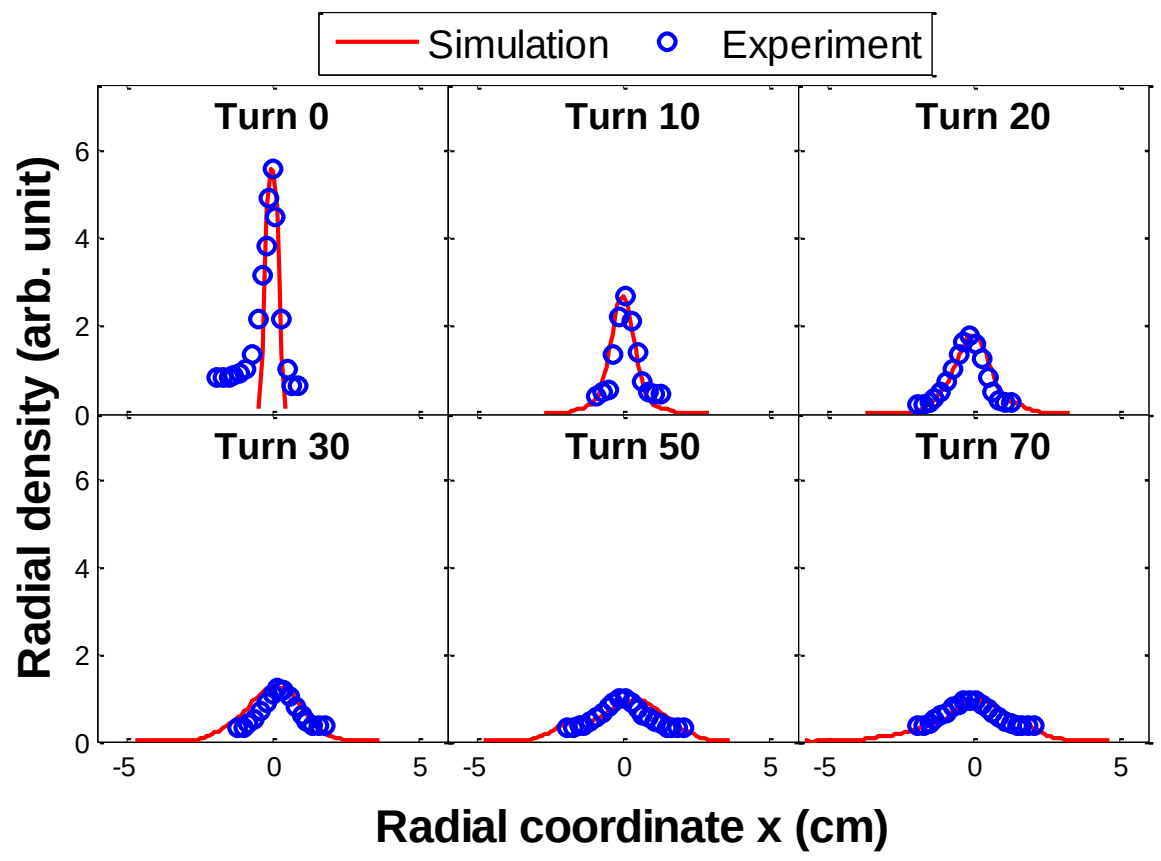

Fig. 14. Evolutions of the radial beam density.

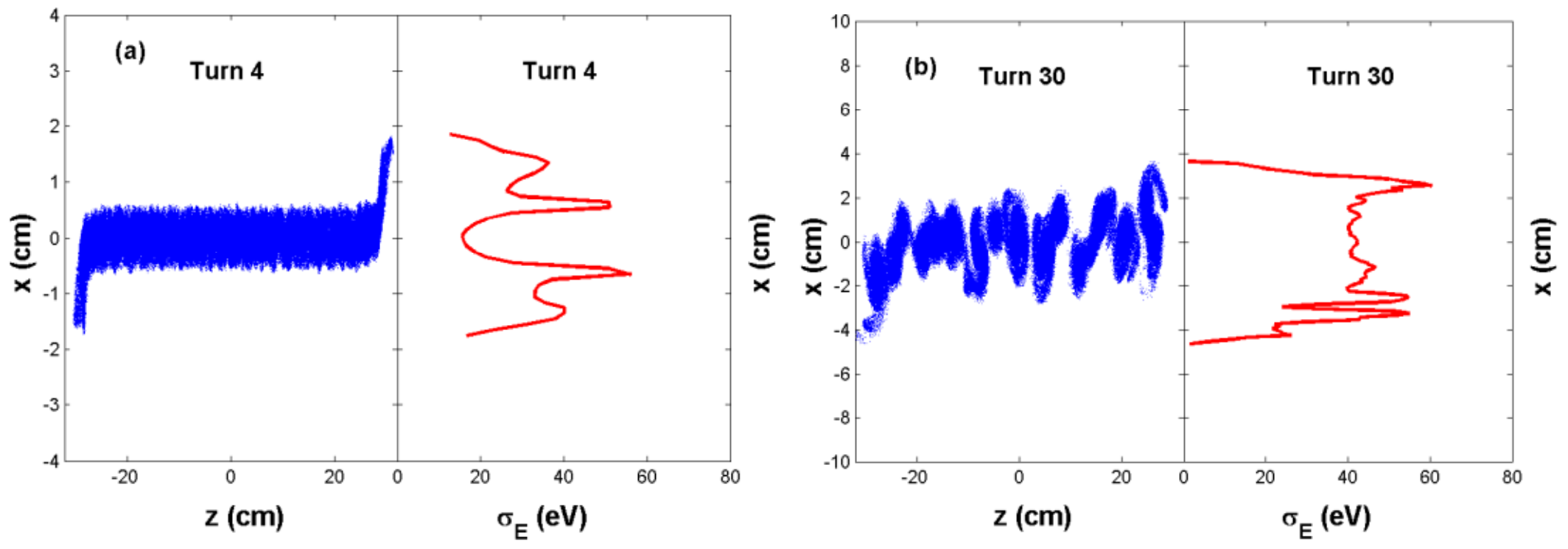

Fig. 15. Simulated top views and RMS slice energy spread at (a) turn 4 and (b) turn 30.

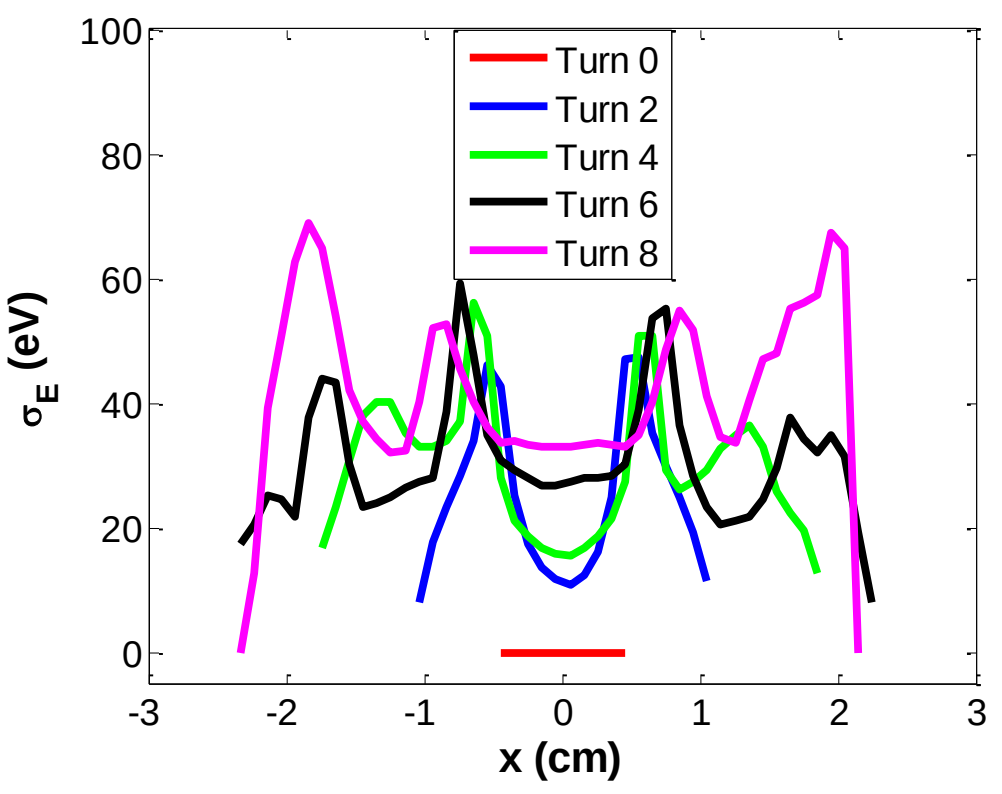

Fig. 16. Simulated RMS slice energy spread at turns 0-8.

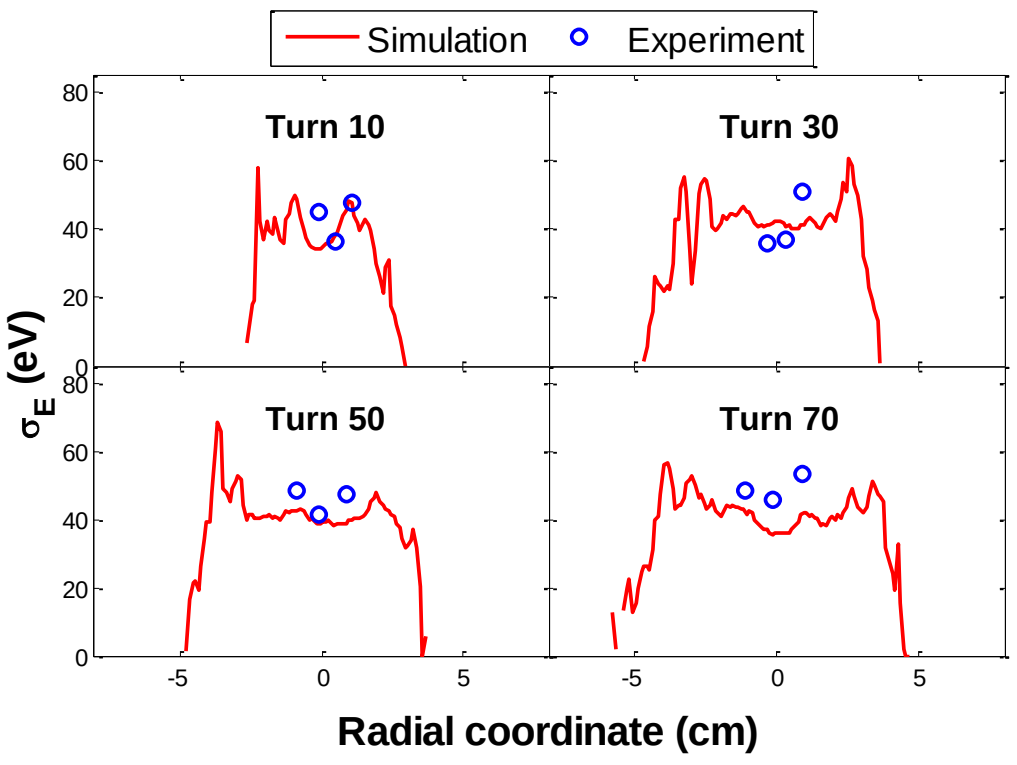

Fig. 17. Comparisons of RMS slice energy spread.

Figs. 15 - 17 show that the space charge fields induce the longitudinal density modulations and energy spread in an initially mono-energetic and straight long coasting bunch in the isochronous ring. At smaller turn numbers, the energy spread in the beam head and tail is much greater than that of the beam core around the beam axis. As the turn number increases, the radial RMS slice energy spread distribution tends to become uniform and changes slowly. At the same time, the radial beam size increases, and the beam centroids deviate from the design orbit. The beam centroid wiggles may also cause the differences in the betatron oscillation phases between the beam clusters (slices). If the beam is long enough, the distribution of the radial centroid offsets of different clusters (slices) may be regarded as randomly uniform around the design orbit. The measured RMS slice energy spread at different radial coordinates is the density-weighted mean RMS slice energy spread of the beam core of any individual cluster (slice) and is independent of the radial coordinates. This can be explained below in Fig. 18.

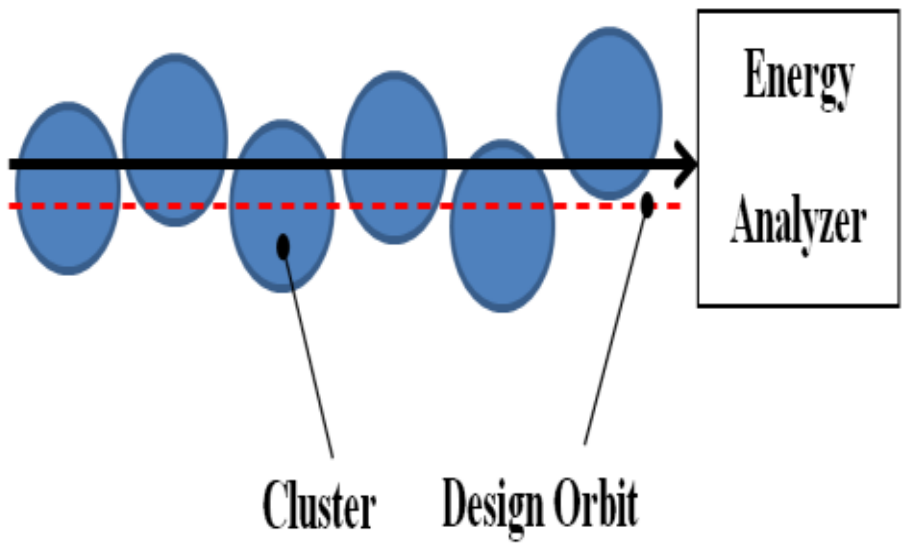

Fig. 18. Sketch of clusters and energy analyzer.

Fig. 18 shows at a given large turn number, the long bunch has broken up into many identical small clusters (the blue ovals) whose centroids distribute randomly and uniformly around the design orbit (the red dashed line). If we measure the RMS slice energy spread at a radial position as indicated by the solid black line with arrow, the analyzer will sample slices of different clusters. Assume there are $N_c$ clusters in the whole bunch that are tagged by ID# *1, 2, 3 ,....*$N_c$, and each cluster has the same number of particles and radial charge distribution profiles. If the slice sampled by the analyzer in each cluster contains $n_i$ ($i = 1, 2, 3,....N_c$) charged particles and its RMS energy spread is $\sigma_i$, the mean kinetic energy of all slices is the same as $< E(x) >$ at large turn numbers, where $x$ is the radial coordinate of the black solid line with respect to the red dashed line. Then the RMS energy spread in the $i^{th}$ beam slice is:

$$\sigma_i = \sqrt{\frac{1}{n_i}\sum_{j=1}^{n_i}[E_j - < E(x) >]^2} \quad , \qquad (i=1,\ 2,\ 3,\ldots\ldots\ N_c\ ). \tag{59}$$

The sum of square of Eq. (59) gives:

$$\frac{n_1\sigma_1^{\ 2} + n_2\sigma_2^2 + \ldots\ldots\ldots n_{N_c}\sigma_{N_c}^2}{n_1 + n_2 + \ldots\ldots n_{N_c}} = < E_i^{\ 2} > - < E(x) >^2 = \sigma_E^2(x), \tag{60}$$

$$\sigma_E(x) = [\frac{n_1\sigma_1^2 + n_2\sigma_2^2 + .........n_{N_c}\sigma_{N_c}^2}{n_1 + n_2 + ......n_{N_c}}]^{\frac{1}{2}}. \tag{61}$$

The RHS of Eq. (61) is the density-weighted mean RMS energy spread of the sampled beam slices of different cluster cores at a fixed radial coordinate *x*. It is actually equal to the density-weighed mean RMS slice energy spread of any given single cluster core and is independent of the coordinate *x*, if the number of clusters is large enough and the radial centroid offsets of all the clusters are randomly and uniformly distributed around the design orbit. In real measurements, the above ideal preconditions are not satisfied completely, hence there are always small energy spread fluctuations among different radial measurement points.

The equilibrium value of the kinetic energy deviation $\Delta E_{eq}(x)=E_{eq}(x)-E_{k0}$ and the radial coordinate *x* of an off-momentum particle satisfy the relation

$$\Delta E_{eq}(x) = \frac{2E_{k0}}{R}x, \tag{62}$$

where $E_{eq}(x)$ is the equilibrium kinetic energy and is equal to $<E(x)>$ of the beam slices centered at *x* at large turn numbers. For simplicity, assume the radial beam density distribution is uniform, then the RMS energy spread of the equilibrium particles measured by the SIR energy analyzer with an entrance slit of width $\Gamma$ = 1 mm centered at *x* can be estimated as:

$$\sigma_{\Delta E_{eq}} = [\frac{1}{\Gamma}\int_{x-\frac{\Gamma}{2}}^{x+\frac{\Gamma}{2}}[\frac{2E_{k0}}{R}(x'-x)]^2 dx']^{\frac{1}{2}} = \frac{E_{k0}\Gamma}{\sqrt{3}R} \approx 5.7eV. \tag{63}$$

This value is proportional to the slit width $\Gamma$ and is independent of *x*. In addition, it is much less than the asymptotic energy spread of about 50 eV at large turn numbers. This indicates that the number of particles at equilibrium energy only accounts for a small fraction of the total particles in a beam slice.

The saturation of the RMS slice energy spread of clusters in the SIR is an indication of formation of the nonlinear advection of the beam in the $\boldsymbol{E} \times \boldsymbol{B}$ velocity field [20]. Assume an ideal disk-shaped cluster coasts in an isochronous ring with an effective uniform magnetic field $\boldsymbol{B}_{eff}$, the $\boldsymbol{E}_{sc} \times \boldsymbol{B}_{eff}$ velocity field at any point on the median plane inside the cluster is along the azimuthal direction in the rest frame of the cluster. This will result in no particles staying at the beam head (tail) forever, accordingly, the energy spread within a given beam slice of 1-mm width at any coordinate *x* will not build up with time significantly. During the binary cluster merging process, the total charge and size of the new clusters grow at the same time. Hence, the mean charge density does not change considerably which may result in the saturation of the mean RMS slice energy spread averaged over the radial coordinate.

## 7. Conclusions

In this paper, we introduced a modified 2D dispersion relation to discuss the Landau damping effects for a coasting beam in the isochronous regime. The radial-longitudinal coupling transfer matrix elements $R_{51}$ and $R_{52}$ are included in the 2D dispersion relation. These elements can modify the coherent slip factor, together with $R_{56}$, they also provide an exponential suppression for the instability growth rates of a beam with finite energy spread and emittance, namely, the Landau damping effects. The 2D dispersion relation is benchmarked by simulation code CYCO for bunches with variable initial beam intensities, energy spread and emittance. The theory agrees well with the numerical simulations for perturbation wavelengths of $\lambda >$ 2.0 cm. While for the cases of $\lambda <$ 2.0 cm, the discrepancies between simulations and theoretical predictions become larger. The possible reasons for the discrepancies are pointed out and discussed. By comparisons, the 2D dispersion relation has an overall better performance than the conventional 1D growth rate formula, which significantly overestimates the growth rates in the short–wavelength limit $\lambda \to 0$ and is incapable of predicting the correct fastest-growing wavelength. The Landau damping effect is a necessary and important mechanism for an accurate prediction of the instability growth rates of the short-wavelength perturbations and the fastest-growing wavelength.

The measured RMS slice energy spread and radial density profiles of a long coasting bunch agree with the simulations. At large turn numbers, the randomly distributed cluster centroid offsets tend to make the radial energy spread distribution of the whole bunch uniform. The measured energy spread is the density-weighted mean RMS slice energy spread of any single cluster core, and its saturation behavior indicates the formation of the nonlinear advection of the particles due to the $\boldsymbol{E} \times \boldsymbol{B}$ velocity field in each cluster.

## Acknowledgements

We would like to thank Prof. F. Marti and T. P. Wangler for their guidance. We are also grateful to Y.C. Wang, S. Y. Lee, K. Y. Ng, G. Stupakov, E. Pozdeyev, A. W. Chao, R. York, M. Syphers, V. Zelevinsky, J. Baldwin, and J. A. Rodriguez for their fruitful discussions and suggestions. This work was supported by NSF Grant # PHY 0606007.